\documentclass[10pt,journal]{IEEEtran}

\usepackage[T1]{fontenc}
\usepackage{cite}
\usepackage{amsmath,graphicx}
\usepackage{amssymb}
\usepackage{algpseudocode}
\usepackage{amsfonts}
\usepackage{graphicx,subfigure}
\usepackage{fancyhdr}  %
\usepackage{cases}
\usepackage{extarrows}
\usepackage{algorithm}
\usepackage{multirow,tabularx}
\usepackage{mathtools}
\usepackage{xcolor}
\usepackage[english]{babel}

\newtheorem{lemma}{Lemma}

\ifCLASSINFOpdf
\else
\fi
\begin{document}
\title{ Reconfigurable Intelligent Surfaces for\\ Energy Efficiency in Wireless Communication}
\author{Chongwen~Huang, Alessio~Zappone,~\IEEEmembership{Senior Member,~IEEE,} George~C.~Alexandropoulos,~\IEEEmembership{Senior Member,~IEEE,} M\'{e}rouane~Debbah,~\IEEEmembership{Fellow,~IEEE,} and Chau~Yuen,~\IEEEmembership{Senior Member,~IEEE}
\thanks{Part of this work has been presented in \textit{IEEE ICASSP}, Calgary, Canada, 14--20 April 2018 \cite{chongwen2018}.}
\thanks{C.~Huang and C. Yuen are with the Singapore University of Technology and Design, 487372 Singapore. (emails: chongwen\_huang@mymail.sutd.edu.sg, yuenchau@sutd.edu.sg)}
\thanks{A.~Zappone and M.~Debbah are with CentraleSup\'elec, University Paris-Saclay, 91192 Gif-sur-Yvette, France. M. Debbah is also with the Mathematical and Algorithmic Sciences Lab, Paris Research Center, Huawei Technologies France SASU, 92100 Boulogne-Billancourt, France. (emails: alessio.zappone@l2s.centralesupelec.fr, merouane.debbah@huawei.com)}
\thanks{G.~C.~Alexandropoulos was with the Mathematical and Algorithmic Sciences Lab, Paris Research Center, Huawei Technologies France SASU, 92100 Boulogne-Billancourt, France. He is now with the Department of Informatics and Telecommunications, National and Kapodistrian University of Athens, Panepistimiopolis Ilissia, 15784 Athens, Greece. (e-mail: alexandg@di.uoa.gr)}}

\maketitle

\begin{abstract}
The adoption of a Reconfigurable Intelligent Surface (RIS) for downlink multi-user communication from a multi-antenna base station is investigated in this paper. We develop energy-efficient designs for both the transmit power allocation and the phase shifts of the surface reflecting elements, subject to individual link budget guarantees for the mobile users. This leads to non-convex design optimization problems for which to tackle we propose two computationally affordable approaches, capitalizing on alternating maximization, gradient descent search, and sequential fractional programming. Specifically, one algorithm employs gradient descent for obtaining the RIS phase coefficients, and fractional programming for optimal transmit power allocation. Instead, the second algorithm employs sequential fractional programming for the optimization of the RIS phase shifts. In addition, a realistic power consumption model for RIS-based systems is presented, and the performance of the proposed methods is analyzed in a realistic outdoor environment. In particular, our results show that the proposed RIS-based resource allocation methods are able to provide up to $300\%$ higher energy efficiency, in comparison with the use of regular multi-antenna amplify-and-forward relaying.
\end{abstract}



\section{Introduction}\label{sec:intro}
The highly demanding data rate requirements of emerging and future wireless networks ($5$-th Generation (5G) and beyond) have raised serious concerns on their energy consumption \cite{5GNGMN,Green_5G_wuqingqing}. These networks are anticipated to connect over $50$ billions of wireless capability devices by 2020 \cite{EricssonWP} via dense deployments of multi-antenna base stations and access points \cite{mmwave_5G,Alexandropoulos_CB,Zuojun_EE}. As a consequence, the bit-per-Joule Energy Efficiency (EE) has emerged as a key performance indicator to ensure green and sustainable wireless networks\cite{5GNGMN,Yuen_EE,Green_5G_wuqingqing}, and several energy efficient wireless solutions have been proposed. A survey on the different approaches to implement energy efficient 5G wireless networks has recently appeared in \cite{GEJSAC16}. Therein, the authors conclude that the energy challenge can be conquered only by the joint use of multiple approaches ranging from the use of renewable energy sources, energy efficient hardware components and relevant deployment techniques, as well as green resource allocation and transceiver signal processing algorithms. The issue of radio resource allocation for EE maximization in wireless networks is addressed in detail in \cite{ZapNow15}, where the related mathematical tools are discussed. In \cite{scaling_up,phaseshifter_constraits,emil_setting,sha_hu} it is established that deploying a massive number of antennas can bring substantial energy-efficient benefits.

Among the candidate transceiver approaches \cite{Reconfigurable_arrays,alexandg_ESPARs,Ralf_LMA,LIS_lowresolution} for green communication, a recent emerging hardware technology with increased potential for significant energy consumption reductions is the so-called reconfigurable intelligent surface (RIS)\cite{sha_hu,hu2018wcnc,wu2018LIS,Liaskos_metasurface,han2019,metasufaces_2018nuture}. A RIS is a meta-surface equipped with integrated electronic circuits that can be programmed to alter an incoming electromagnetic field in a customizable way. It consists of a single- or few-layer stack of planar structures that can be readily fabricated using lithography and nano-printing methods. Each RIS unit is implemented by reflect-arrays that employ varactor diodes or other Micro-Electrical-Mechanical Systems, and whose resonant frequency is electronically controlled \cite{Reconfigurable_arrays,Chen2016,Liaskos_metasurface,Liaskos_COMMMAG,tan_indoor,tan_infocom2018,Foo_LC}. The RIS units operating on the incoming field can be distributed over the meta-surface with continuity \cite{Hu_2018,wu2018LIS,sha_hu,hu2018wcnc} or in discrete positions \cite{han2019,metasufaces_2018nuture,Reconfigurable_arrays,Yang_metasurfaces_2016}.

Regardless of the specific implementation, what makes the RIS technology attractive from an energy consumption standpoint, is the possibility of amplifying and forwarding the incoming signal without employing any power amplifier, but rather by suitably designing the phase shifts applied by each reflecting element, in order to constructively combine each reflected signal. Clearly, since no amplifier is used, a RIS will consume much less energy than a regular Amplify-and-Forward (AF) relay transceiver. Moreover, RIS structures can be easily integrated in the communication environment, since their very low hardware footprint allows for their easy deployment into buildings facades, room and factory ceilings, laptop cases, up to being integrated into human clothing \cite{Heath_DAS,LIS_lowresolution}. On the other hand, the lack of an amplifier implies that the gain of a RIS will be lower than what can be achieved by a traditional AF relay with a number of antennas equal to the number of reflecting elements of the RIS. Thus, as far as EE is concerned, it is not clear if a RIS-based system is more convenient than traditional AF relay-based systems \cite{relay_model,Relay_precoding,relay_heath,relay_allessio}. This work aims at answering this question, showing that by properly designing the phase shifts applied by the RIS leads to higher EE than that of a  communication system based on traditional  AF relays.

Available works on RIS-based communication system have mainly focused on the rate performance of RIS in indoor environments. In \cite{Kaina_metasurfaces_2014}, a $0.4m^2$ and $1.5mm$ thickness planar meta-surface consisting of $102$ controllable electromagnetic unit cells and operating at $2.47$GHz was designed. Each unit cell is a rectangular patch sitting on a ground plane \cite{Kaina_resonant_2014} and offering binary phase modulation. The meta-surface was deployed as a spatial microwave modulator in a typical office room, and was demonstrated that it can passively increase the received signal power by an order of magnitude, or completely null it. Integrating each unit cell with one PIN diode in \cite{Yang_metasurfaces_2016}, a programmable meta-surface with dynamic polarization, scattering, and focusing control was designed. The presented experiments demonstrated various controllable electromagnetic phenomena, including anomalous reflection and diffusion, as well as beamforming\cite{wu2018LIS,wu2018lisglobecom}. A detailed analysis on the information transfer from multiple users to a RIS with active elements was carried out in \cite{sha_hu}, while \cite{tan_indoor} showed that a reflectarray can effectively achieve amplification gains, while at the same time cancel interference. In this work, the results were also corroborated by measurements carried out on an office testbed. The role of RIS in improving indoor coverage was also analyzed in \cite{Subrt_control,Subrt_control01}. Very recently, \cite{tan_infocom2018} experimented on the incorporation of a smart reflectarray in an IEEE 802.11ad network operating in the unlicensed $60$GHz frequency band. In \cite{Liaskos_metasurface} the adoption of intelligent meta-surfaces for actively reprogramming communication environments is envisioned, and its advantages in terms of coverage, energy saving, and security are discussed. 

In this work, we consider the downlink of an outdoor cellular network in which a multiple-antenna BS reaches the single-antenna mobile users through a RIS that forwards a suitably phase-shifted version of the transmitted signal. A discrete RIS implementation is adopted, in which a finite number of reflecting units are equipped on the intelligent surface. The contributions of this paper are summarized as follows.
\begin{itemize}
	\item We develop a realistic RIS power consumption model that is based on the number of deployed reflector units and their phase resolution capability. This allows us to formulate the EE maximization problem to optimize the RIS phase shifts and the downlink transmit powers under maximum power and minimum Quality of Service (QoS) constraints.
\item We develop two novel low complexity, and provably convergent, optimization algorithms to tackle the EE maximization problem that account for the unit-modulus constraint of the RIS phase shifts, that is not to be enforced in the design of traditional relay systems. The proposed algorithms exploit the frameworks of alternating maximization, sequential fractional programming, and conjugate gradient search.
  \item We numerically evaluate the performance of the proposed algorithms in a realistic outdoor scenario, while all previous works about RIS have focused on indoor settings. Our results indicate that the proposed algorithms are indeed able to grant higher EE performance than that obtained by a traditional relay-assisted communication system.
\end{itemize}

The remainder of this paper is organized as follows. In Section \ref{sec:format}, the multi-user MISO system assisted by reconfigurable RIS structures is described and the targeted EE maximization problem is formulated. The two derived algorithms for the EE maximization design are presented in Section \ref{sec:EE}, and extensive numerically evaluated results are provided in Section \ref{Sec:Numerics}. Finally, concluding remarks and future research directions are drawn in Section~\ref{sec:prior}.

\textit{Notation}: $a$ is a scalar, $\mathbf{a}$ is a vector, and $\mathbf{A}$  is a matrix. $\mathbf{A}^T$, $\mathbf{A}^H$, $\mathbf{A}^{-1}$, $\mathbf{A^+}$, and $\|\mathbf{A}\|_F$ denote transpose, Hermitian (conjugate transpose), inverse, pseudo-inverse, and Frobenius norm of $ \mathbf{A} $, respectively. $\mathrm{Re}(\cdot)$, $\mathrm{Im}(\cdot)$, $|\cdot|$, $(\cdot)^*$ and $\mathrm{arg}(\cdot)$ denote the real part, imaginary part, modulus, conjugate and the angle of a complex number, respectively. $\text{tr}(\cdot)$ denotes the trace of a matrix and $\mathbf{I}_n$ (with $n\geq2$) is the $n\times n$ identity matrix. $ \mathbf{A} \circ \mathbf{B} $ and $ \mathbf{A} \otimes \mathbf{B} $ denote the Hadamard and Kronecker products of $\mathbf{A}$ and $\mathbf{B}$, respectively, while $\text{vec}(\mathbf{A})$ is a vector stacking all the columns of $\mathbf{A}$. $ \mathrm{diag}(\mathbf{a})$ is a diagonal matrix with the entries of $\mathbf{a}$ on its main diagonal. $\mathbf{A}\succeq\mathbf{B} $ means that $\mathbf{A}-\mathbf{B}$ is positive semidefinite. Notation $x\sim\mathcal{CN}(0,\sigma^2)$ means that random variable $x$ is complex circularly symmetric Gaussian with zero mean and variance $\sigma^2$, whereas $E[x]$ denotes $x$'s expected value. $\mathbb{R}$ and $\mathbb{C}$ denote the complex and real number sets, respectively, and $j\triangleq\sqrt{-1}$ is the imaginary unit.

\section{System Model}\label{sec:format}
In this section, we describe the signal model for the considered RIS-based downlink multi-user MISO system, and then develop a model for the system total power consumption, which accounts for the energy consumption at the RIS. This section also describes the problem formulation for the joint design of the  users' transmit powers and RIS phase shifts.
\subsection{Signal Model}
\begin{figure} \vspace{-1mm}
  \begin{center}
  \includegraphics[width=95mm]{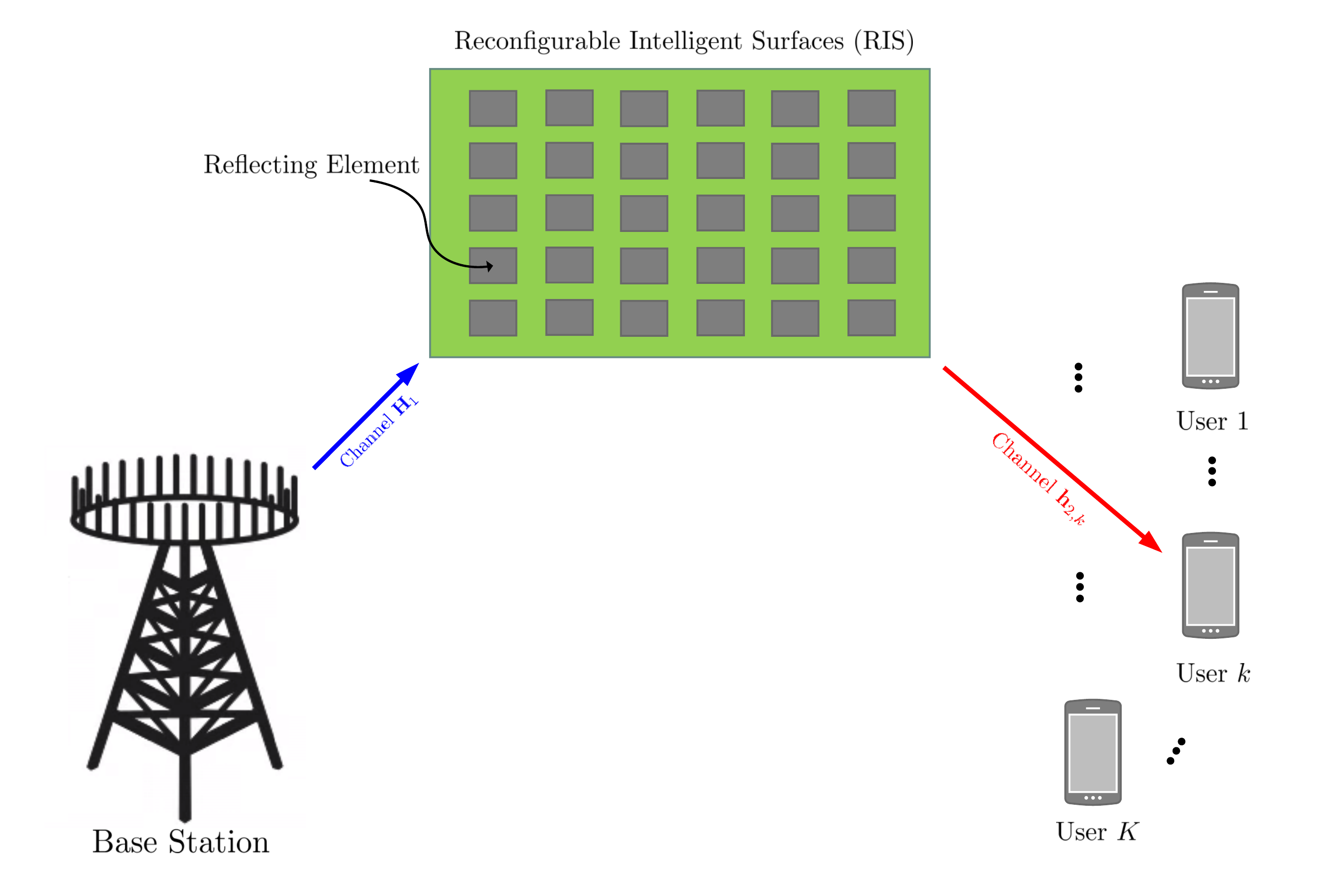}  \vspace{-2mm}
  \caption{The considered RIS-based multi-user MISO system comprising of a $M$-antenna base station simultaneously serving in the downlink $K$ single-antenna users. RIS is assumed to be attached to a surrounding building's facade, and the transmit signal propagates to the users via the assistance of RIS that is capable of reconfigurable behavior.}
  \label{fig:Estimation_Scheme} \vspace{-6mm}
  \end{center}
\end{figure}

Consider the downlink communication between a BS equipped with $M$ antenna elements and $K$ single-antenna mobile users. We assume that this communication takes place via a RIS with $N$ reflecting elements deployed on the facade of a building existing in the vicinity of both communication ends, as illustrated in Fig$.$~\ref{fig:Estimation_Scheme}. The direct signal path between the BS and the mobile users is neglected due to unfavorable propagation conditions. 

Then, the discrete-time signal received at mobile user $k$, with $k=1,2,\ldots,K$, is written as
\begin{equation}\label{model_01}
  y_k=
  \mathbf{h}_{2,k}\mathbf{\Phi}\mathbf{H}_{1}\mathbf{x}+w_{k}
\end{equation}
where $\mathbf{h}_{2,k}\in\mathbb{C}^{1\times N}$ denotes the channel vector between the RIS and user $k$, $\mathbf{H}_{1} \in\mathbb{C}^{N\times M}$ denotes the channel matrix between the BS and the RIS, $\mathbf{\Phi}\triangleq\mathrm{diag}[\phi_1,\phi_2,\ldots,\phi_N]$ is a diagonal matrix accounting for the effective phase shifts applied by all RIS reflecting elements, where $\phi_n=e^{j\theta_n}$ $\forall$$n=1,2,\ldots,N$, while $w_{k}\sim\mathcal{CN}(0,\sigma^{2})$ models the thermal noise power at receiver $k$.

Finally, $\mathbf{x}\triangleq\sum_{k=1}^{K}\sqrt{p}_{k}\mathbf{g}_{k}s_{k}$ denotes the transmitted signal with $p_{k}$, $s_{k}$, and $\mathbf{g}_{k}\in\mathbb{C}^{M\times 1}$ representing the transmit power, unit-power complex valued information symbol chosen from a discrete constellation set, and precoding vector, respectively, intended for the $k$-th mobile user.
The power of the transmit signal from the multi-antenna BS is subject to the maximum transmit power constraint:
\begin{equation}\label{model_03}
  E[|\mathbf{x}|^2]=\mathrm{tr}(\mathbf{P}\mathbf{G}^H\mathbf{G})\leq P_{\rm max}\;,
\end{equation}
wherein $\mathbf{G}\triangleq[\mathbf{g}_1,\mathbf{g}_2,...,\mathbf{g}_K]\in\mathbb{C}^{M\times K}$, $\mathbf{P}\triangleq\mathrm{diag}[p_1,...,p_K]\in\mathbb{R}^{K\times K}$.

As seen from \eqref{model_01}, the reflecting surface is modeled as a scatterer with reconfigurable characteristics. It effectively applies the phase shifting operation described by $\mathbf{\Phi}$ to the impinging information bearing signal expressed by $\mathbf{H}_{1}\mathbf{x}$. Thus, a RIS operates in way that resembles an AF relay, with the crucial difference that no power amplifier is present in a RIS. As a result, unlike what happens in AF relaying, each RIS coefficient is constrained to have unit modulus. Moreover, a RIS does not perform any decoding, and/or digitalization operation, operating instead directly on the RF incoming signal \cite{Subrt_control,tan_indoor,Subrt_control01,Reconfigurable_arrays,tan_infocom2018}. As a result, a RIS is anticipated to require a much lower energy consumption than a traditional  AF relay, requiring only a limited, static power supply.

Based on \eqref{model_01}, the Signal-to-Interference-plus-Noise Ratio (SINR) experienced at the $k$-th mobile user is obtained as
\begin{equation}\label{model_05}
  \gamma_k\triangleq\frac{p_k|\mathbf{h}_{2,k}\mathbf{\Phi} \mathbf{H}_{1}\mathbf{g}_k|^2}{\sum\limits_{i=1,i\neq k}^{K} p_i|\mathbf{h}_{2,k}\mathbf{\Phi} \mathbf{H}_{1}\mathbf{g}_i|^2+\sigma^2}\;.
\end{equation}
Then, based on \eqref{model_05}, the system Spectral Efficiency (SE) in bps/Hz is given by
\begin{equation}\label{model_07}
  \mathcal{R}\triangleq\sum\limits_{k=1}^{K}\mathrm{log}_2(1+\gamma_k).
\end{equation}

\subsection{Total Power Consumption Model}\label{Sec:PowerConsumption}
The total power dissipated to operate the considered RIS-based system is composed of the BS transmit power, as well as the hardware static power consumed in the BS, mobile user terminals, and RIS. In particular, it should be stressed that the RIS does not consume any transmit power, since its reflectors are passive elements that do not directly alter the magnitude of the incoming signal. Any amplification gain provided by the RIS is obtained through a suitable adjustment of the phase shifts of the reflecting elements so as to recombine the reflected signals with a phase coherence. As previously mentioned, this is a significant difference compared to AF relay architectures \cite{B:Relay_Book}, which instead rely on dedicated power amplifier to achieve amplification gains.

Putting all above together, the power consumption of the $k$-th wireless link between BS and the $k$-th mobile user can be expressed as
\begin{align}\label{power_model1}
   \mathcal{P}_k\triangleq\xi p_k + P_{{\rm UE},k}+ P_{\rm BS} + P_{\rm RIS},
\end{align}
wherein $\xi\triangleq\nu^{-1}$ with $\nu$ being the efficiency of the transmit power amplifier, $P_{{\rm UE},k}$ is the hardware static power dissipated by the $k$-th user equipment, while $P_{\rm BS}$ and $P_{\rm RIS}$ denote the total hardware static power consumption at BS and RIS, respectively. We should remark that the two underlying assumptions in \eqref{power_model1} are: \textit{i}) the transmit amplifier operates in its linear region; and \textit{ii}) the circuit power $P_{\rm c}$ does not depend on the communication rate. Both assumptions are met in typical wireless communication systems, which are operated so as to ensure that the amplifiers operate in the linear region of their transfer function, and in which the hardware-dissipated power can be approximated by a constant power offset.


The RIS power consumption depends on the type and the resolution of its individual reflecting elements that effectively perform phase shifting on the impinging signal. Typical power consumption values of each phase shifter are $1.5$, $4.5$, $6$, and $7.8$mW for $3$-, $4$-, $5$-, and $6$-bit resolution phase shifting \cite{powerconsup_lowadc,phaseshifter_powerconsup,LIS_lowresolution,chogwentsp2019}. Therefore, the power dissipated at an intelligent surface with $N$ identical reflecting elements can be written as $P_{\rm RIS}=NP_{n}(b)$, where $P_{n}(b)$ denotes the power consumption of each phase shifter having $b$-bit resolution. Using the latter definitions, the total amount of power needed to operate the RIS-based downlink multi-user MISO system is given by
\begin{align}\label{power_model3}
   \mathcal{P}_{\rm total}= &\sum_{k=1}^{K}(\xi p_k + P_{{\rm UE},k}) + P_{\rm BS}+ NP_{n}(b).
\end{align}  \vspace{-2mm}

In the following, the power model in \eqref{power_model3} will be considered as the denominator of the energy efficiency function, even though the optimization of the number of quantization bits $b$ will not be considered, and left for thorough investigation in future work. Some preliminary results on the optimization of $b$ have appeared in \cite{LIS_lowresolution}.


\subsection{Design Problem Formulation} \label{sec:problem}
We are interested in the joint design of the transmit powers for all users, included in $\mathbf{P}=\mathrm{diag}[p_1,p_2,\ldots,p_K]$, and the values for the RIS elements, appearing in the diagonal of $\mathbf{\Phi}=\mathrm{diag}[\phi_1,\phi_2,\ldots,\phi_N]$, that jointly maximize the bit-per-Joule EE performance for the considered RIS-based system. This performance is defined as the ratio between the system achievable sum rate in bps and the total power consumption in Joule, i$.$e$.$, $\eta_{\rm EE}\triangleq{\rm BW}\mathcal{R}/\mathcal{P}_{\rm total}$ with ${\rm BW}$ being the transmission bandwidth, and can be obtained using \eqref{model_07} and \eqref{power_model3} as
\begin{align} \label{powerEE}
&\displaystyle
\eta_{\rm EE} =\frac{{\rm BW}\sum_{k=1}^{K} \mathrm{log_2}(1+\gamma_k)}{\xi\sum_{k=1}^{K}p_k + P_{\rm BS} + KP_{\rm UE} + NP_{n}(b)}.
\end{align}
The EE maximization will be carried out enforcing maximum power constraints as well as individual QoS requirements for all $K$ mobile users. To make the targeted problem more tractable, we assume that: \textit{i}) the reflectors phase shifting resolution is infinite, i$.$e$.$, $2^{b}>>1$; and \textit{ii}) all involved channels are perfectly known at BS that employs Zero-Forcing (ZF) transmission, which is known to be optimal in the high-SINR regime\footnote{It is well-known that in general the optimal linear receive structure is the LMMSE filter. However, this has the drawback of not suppressing the multi-user interference, which would significantly complicate the resource allocation phase.} \cite{DE_ZFR,chongwenTVT}. To this end, it is assumed that the BS perfectly knows all communication channels $\mathbf{h}_{2,k}$ $\forall$$k=1,2,\ldots,K$ and $\mathbf{H}_1$, which can be acquired by the methods   described in e.g. \cite{Alexandropoulos_position, Alkhateeb_DNN}.

Assuming that the equivalent channel matrix $\mathbf{H}_{2}\mathbf{\Phi} \mathbf{H}_1$ has a right inverse\footnote{Note that this happens with probability one if, for example, $M\geq K=N$, i.e. if the RIS activates a number of reflecting elements equal to the number of users to serve. Activating more reflectors would lead to higher energy consumptions.},  perfect interference suppression is achieved by setting the ZF precoding matrix to $\mathbf{G}=(\mathbf{H}_{2}\mathbf{\Phi} \mathbf{H}_1)^{+}$, where $\mathbf{H}_2\triangleq[\mathbf{h}_{2,1}^T,\mathbf{h}_{2,2}^T,\ldots,\mathbf{h}_{2,K}^T]^T\in\mathbb{C}^{K\times N}$.

Substituting this $\mathbf{G}$ in \eqref{model_05}, the considered EE maximization problem is expressed as follows:
\begin{subequations}\label{Prob:ResAllpower}
\begin{align}
&\displaystyle \max_{\mathbf{\Phi},\mathbf{P}}\;\frac{\sum_{k=1}^{K}\log_2\left(1+p_k\sigma^{-2}\right)}{\xi\sum_{k=1}^{K}p_k+P_{\rm BS}+KP_{\rm UE}+NP_{n}(b) }\label{Prob:aResAllpower}\\
&\;\text{s.t.}\;\log_2\left(1+p_k\sigma^{-2}\right)\geq R_{{\rm min},k}\;\forall k=1,2,\ldots,K,\label{Prob:bProb:ResAllpower}\\
&\;\quad\;\; \text{tr}((\mathbf{H}_{2}\mathbf{\Phi} \mathbf{H}_{1})^{+}\mathbf{P}(\mathbf{H}_{2}\mathbf{\Phi} \mathbf{H}_{1})^{+H})\leq P_{\max}, \label{Prob:cResAllpower}\\
&\;\quad\;\; |\phi_n|=1\;\forall n=1,2,\ldots,N, \label{Prob:dResAllpower}
\end{align}
\end{subequations}
where $R_{{\rm min},k}$ denotes the individual QoS constraint of the $k$-th user. Also, constraint \eqref{Prob:cResAllpower} ensures that the BS transmit power is kept below the maximum feasible threshold $P_{\max}$, while constraint \eqref{Prob:dResAllpower} accounts for the fact that each RIS reflecting element can only provide a phase shift, without amplifying the incoming signal.

The optimization problem \eqref{Prob:ResAllpower} is non-convex, and is made particularly challenging by  the presence of $\mathbf{\Phi}$. In the sequel, we present two computationally efficient approaches to tackle \eqref{Prob:ResAllpower}.

\section{Energy Efficiency Maximization} \label{sec:EE}
Solving the optimization problem described in \eqref{Prob:ResAllpower} is challenging mainly due to the constraints \eqref{Prob:cResAllpower} and \eqref{Prob:dResAllpower}. In order to develop a tractable algorithm for the design paramateres, a convenient approach is to employ the alternating optimization technique \cite{J:alternating_minimization} to separately and iteratively solve for $\mathbf{P}$ and $\mathbf{\Phi}$. In particular, we first solve for $\mathbf{\Phi}$ given a fixed $\mathbf{P}$, and then find the optimum $\mathbf{P}$ when $\mathbf{\Phi}$ is fixed. Iterating this process improves the EE value at each iteration step, and must eventually converge in the optimum value of the objective, since \eqref{Prob:aResAllpower} is upper-bounded on the feasible set. In the rest of this section, the optimization with respect to $\mathbf{\Phi}$ for fixed $\mathbf{P}$, and with respect to $\mathbf{P}$ for fixed $\mathbf{\Phi}$ will be treated separately.

\subsection{Optimization with respect to the RIS Elements Values $\mathbf{\Phi}$}\label{sec:respect_Phi}
For a fixed transmit power allocation matrix $\mathbf{P}$, the design problem \eqref{Prob:ResAllpower} becomes the following feasibility test:
\begin{subequations}\label{Eq:ResAllProbPhi0}
\begin{align}
&\displaystyle \max_{\mathbf{\Phi}}\;C_{o} \\
&\;\text{s.t.}\;\text{tr}((\mathbf{H}_{2}\mathbf{\Phi} \mathbf{H}_{1})^{+}\mathbf{P}(\mathbf{H}_{2}\mathbf{\Phi} \mathbf{H}_{1})^{+H})\leq P_{\rm max}, \label{Eq:bResAllProbPhi0}\\
&\;\quad\;\;|\phi_n|=1\;\forall n=1,2,\ldots,N, \label{Eq:cResAllProbPhi0}
\end{align}
\end{subequations}
wherein $C_{o}$ denotes any constant number. Since the power allocation matrix $\mathbf{P}$ is fixed, the objective can be reduced as a constant-value objective function $C_{o}$ under the constraints  \eqref{Eq:bResAllProbPhi0} and \eqref{Eq:cResAllProbPhi0}. The challenge in solving problem \eqref{Eq:ResAllProbPhi0} lies in the fact that its objective is non-differentiable and that the feasible set is not convex. To proceed further, we observe that \eqref{Eq:ResAllProbPhi0} is feasible if and only if the solution of the following optimization problem:
\begin{subequations}\label{Eq:ResAllProbPhi1}
\begin{align}
&\displaystyle \min_{\mathbf{\Phi} }\text{tr}((\mathbf{H}_{2}\mathbf{\Phi} \mathbf{H}_{1})^{+}\mathbf{P}(\mathbf{H}_{2}\mathbf{\Phi} \mathbf{H}_{1})^{+H}) \label{Eq:aResAllProbPhi1}\\
&\;\text{s.t.}\;|\phi_n|=1\;\forall n=1,2,\ldots,N \label{Eq:bResAllProbPhi1}
\end{align}
\end{subequations}
is such that the objective can be made lower than $P_{\rm max}$.

Let us define $\mathbf{\Theta}\triangleq\mathrm{diag}[\theta_{1},\theta_{2},\ldots,\theta_{N}]$ and then express $\mathbf{\Phi}$ as the following function of $\mathbf{\Theta}$ (recall that $\phi_n=e^{j\theta_n}$ $\forall$$n=1,2,\ldots,N$): $\mathbf{\Phi}(\mathbf{\Theta})=\mathrm{diag}[e^{j\theta_{1}},e^{j\theta_{2}},\ldots,e^{j\theta_{N}}]$. Using these definitions, problem \eqref{Eq:ResAllProbPhi1} can be reformulated as the following unconstrained problem:
\begin{equation}\label{Prob:Equalpower1}
\displaystyle \min_{\mathbf{\Theta}}\;\mathcal{F}(\mathbf{\Phi}(\mathbf{\Theta}))\triangleq\text{tr}((\mathbf{H}_{2}\mathbf{\Phi(\Theta)} \mathbf{H}_{1})^{+}\mathbf{P}(\mathbf{H}_{2}\mathbf{\Phi}(\mathbf{\Theta}) \mathbf{H}_{1})^{+H}).
\end{equation}
By expressing the power allocation matrix $\mathbf{P}$ as $\mathbf{P}=\mathbf{Q}\mathbf{Q}^T$ with $\mathbf{Q}\triangleq\sqrt{\mathbf{P}}$, we observe that the objective function in \eqref{Prob:Equalpower1} can be rewritten as
\begin{align}
&\mathcal{F}(\mathbf{\Phi}(\mathbf{\Theta}))=\text{tr}((\mathbf{H}_{2}\mathbf{\Phi}(\mathbf{\Theta})\mathbf{H}_{1})^{+}\mathbf{P}(\mathbf{H}_{2}\mathbf{\Phi}(\mathbf{\Theta})\mathbf{H}_{1})^{+H}) \nonumber \\
&=\text{tr}((\mathbf{Q}^{-1}\mathbf{H}_{2}\mathbf{\Phi}(\mathbf{\Theta})\mathbf{H}_{1})^{+}(\mathbf{Q}^{-1}\mathbf{H}_{2}\mathbf{\Phi}(\mathbf{\Theta})\mathbf{H}_{1})^{+H}) \nonumber\\
&\overset{a}{=}\text{tr}((\mathbf{\overline{H}}_{2}\mathbf{\Phi}(\mathbf{\Theta})\mathbf{H}_{1})^{+}(\mathbf{\overline{H}}_{2}\mathbf{\Phi}(\mathbf{\Theta})\mathbf{H}_{1})^{+H}) \nonumber\\
&\overset{b}{=}\|\mathbf{H}_{1}^{+}\mathbf{\Phi}^{-1}(\mathbf{\Theta})\mathbf{\overline{H}}_{2}^{+}\|^2_{F}\overset{c}{=}\|\text{vec}(\mathbf{H}_{1}^{+}\mathbf{\Phi}^{-1}(\mathbf{\Theta})\mathbf{\overline{H}}_{2}^{+})\|^2 \nonumber\\
&=\|(\mathbf{\overline{H}}_{2}^{+H}\!\otimes\mathbf{H}_{1}^{+})\text{vec}(\mathbf{\Phi}^{-1}(\mathbf{\Theta}))\|^2\; \nonumber \\
&=\text{vec}(\mathbf{\Phi}^{-1}(\mathbf{\Theta}))^H(\mathbf{\overline{H}}_{2}^{+H}\!\!\otimes\!\mathbf{H}_{1}^{+})^H(\mathbf{\overline{H}}_{2}^{+H}\!\!\otimes\!\mathbf{H}_{1}^{+})\text{vec}(\mathbf{\Phi}^{-1}(\mathbf{\Theta})). \label{Eq:eObjResAllProbPhi2}
\end{align}
In the latter expression, we have used the definition $\mathbf{\overline{H}}_{2}\triangleq\mathbf{Q}^{-1}\mathbf{H}_{2}$ in step $(a)$,
whereas steps $(b)$ follows from the properties of the Frobenius norm and Pseudo-inverse law of a matrix product,\footnote{A sufficient condition for $(\mathbf{\overline{H}}_{2}\mathbf{\Phi}\mathbf{H}_{1})^{+}=\mathbf{H}_{1}^{+}\mathbf{\Phi}^{-1}\mathbf{\overline{H}}_{2}^{+}$ is that $\mathbf{\overline{H}}_{2}$ has full column rank and $\mathbf{\Phi}(\mathbf{\Theta})\mathbf{H}_{1}$ has full row rank. This happens with probability one if $K \geq N$, $M \geq N$, and recalling that $\mathbf{H}_1$ and $\mathbf{H}_2$ are realizations of random matrices with i.i.d. entries.} and $(c)$ follows from the vectorization operator, respectively. While the alternative expression in \eqref{Eq:eObjResAllProbPhi2} does not lead to a convex problem formulation, it lends itself to being handled by the two efficient approaches to be described in the next two subsections.

\subsubsection{Gradient Descent Approach}
The optimization problem $\min_{\mathbf{\Theta}}\mathcal{F}(\mathbf{\Phi}(\mathbf{\Theta}))$ using \eqref{Eq:eObjResAllProbPhi2} is an unconstrained problem, hence, we can employ gradient search to monotonically decrease its objective, eventually converging to a stationary point. To begin with, we first define the matrices:
\begin{align} \label{Prob:definition}
\mathbf{A}&\triangleq(\mathbf{\overline{H}}_{2}^{+H}\otimes\mathbf{H}_{1}^{+})^H(\mathbf{\overline{H}}_{2}^{+H}\otimes\mathbf{H}_{1}^{+})\in\mathbb{C}^{N^2\times N^2}, \\
\mathbf{y}&\triangleq\mathrm{vec}(\mathbf{\Phi}^{-1}(\mathbf{\Theta}))\in\mathbb{C}^{N^2\times 1}\label{eq:y},
\end{align}
which help us to express \eqref{Eq:eObjResAllProbPhi2} as $\mathcal{F}(\mathbf{\Phi}(\mathbf{\Theta}))=\mathbf{y}^H\mathbf{A}\mathbf{y}$. To compute the gradient of $\mathcal{F}(\mathbf{\Phi}(\mathbf{\Theta}))$ with respect to $\mathbf{\Theta}$, i$.$e$.$, $\nabla_{\mathbf{\Theta}}(\mathbf{y}^H\mathbf{A}\mathbf{y})$, we expand $\mathbf{y}^H\mathbf{A}\mathbf{y}$ using \eqref{eq:y}, which, upon exploiting the fact that $\mathbf{A}=\mathbf{A}^H$, leads to:
\begin{align} \label{Prob:Grad2}
\mathbf{y}^H\!\mathbf{A}\mathbf{y}\!=\!\sum_{n=1}^{N}a_{\ell(n),\ell(n)}+\!2\Re\left\{\sum^{N}_{n=1}\!\sum^{N}_{m>n}a_{\ell(n),\ell(m)}e^{j(\theta_n-\theta_m)}\right\},
\end{align}
where $\ell(n)$ is the index map $\ell(n)=(n-1)N+n$, for all $n=1,\ldots,N$, and $a_{\ell(n),\ell(m)}$ denotes the $(\ell(n),\ell(m))$-th element of $\mathbf{A}$. Then, the derivative of $\mathbf{y}^H\mathbf{A}\mathbf{y}$ with respect to $\theta_i$ $\forall$$i=1,2,\ldots,N$ can be computed as
\begin{align} \label{Prob:Grad3}
\frac{\partial(\mathbf{y}^H\mathbf{A}\mathbf{y})}{\partial \theta_i}=
&2\Re \Bigg\{je^{j\theta_{i}}\sum^{N}_{m>i}a_{\ell(i),\ell(m)}e^{-j\theta_{m}} \nonumber\\
&-je^{-j\theta_{i}}\sum^{N}_{n<i}a_{\ell(n),\ell(i)}e^{j\theta_{n}}\Bigg\}.
\end{align}
Next, a suitable step size for the gradient descent needs to be computed. To this end, denote by $\text{vec}(\mathbf{\Theta})^{t}$ the phase of $\mathbf{y}$ at iteration $t$, and by $\mathbf{d}^{t}$ the adopted descent direction at iteration $t$. Then, the next iteration point is given by
\begin{align}
&\text{vec}(\mathbf{\Theta})^{(t+1)} =\text{vec}(\mathbf{\Theta})^{(t)}+\mu\mathbf{d}^{(t)}, \\
 \label{Prob:Stepsize}
&\mathbf{y}^{(t+1)}= e^{j\text{vec}(\mathbf{\Theta})^{(t+1)}} \circ  \text{vec}(\mathbf{I}_N)=\mathbf{y}^{(t)}
\circ e^{j\mu\mathbf{d}^{(t)}},
\end{align}
where $\mu>0$ is the step size. In order to find a suitable step size, we need to solve the following minimization problem:
\begin{align} \label{Prob:Stepsize1}
\min_{\mu>0}h(\mu)\triangleq\left(\mathbf{y}^{(t+1)}\right)^H\mathbf{A}\mathbf{y}^{(t+1)},
\end{align}
To this end, we expand $h(\mu)$ by plugging \eqref{Prob:Stepsize} into \eqref{Prob:Grad2}, which yields
\begin{align} \label{Prob:Stepsize2}
h(\mu)&= \sum_{n=1}^{N}a_{\ell(n),\ell(n)} \nonumber \\
&+2\Re\left\{\sum^{N}_{n=1}\sum^{N}_{m>n}a_{\ell(n),\ell(m)}e^{j(\theta_{n}^{(t)}-\theta_{m}^{(t)})}e^{j\mu(d_{m}^{(t)}-d_{n}^{(t)})}\right\},
\end{align}
wherein $d_n^{(t)}$ is the $n$-th element of $\mathbf{d}^{(t)}$. Then, $h(\mu)$ can be minimized by means of a line search over $\mu>0$. On the other hand, aiming at a complexity reduction, one can consider a quadratic approximation of \eqref{Prob:Stepsize2} by considering the second-order Taylor expansion of the term $e^{j\mu(d_m^{(t)}-d_n^{(t)})}$ around $\mu=0$, which yields the following approximation of $h(\mu)$
\begin{align}\label{Prob:Stepsizeapp2}
\hat{h}(\mu)=&\sum_{n=1}^{N}a_{\ell(n),\ell(n)}+2\Re\Bigg\{\sum^{N}_{n=1}\sum^{N}_{m>n}a_{\ell(n),\ell(m)}e^{j(\theta_{n}^{(t)}-\theta_{m}^{(t)})} \nonumber \\
&\times\bigg(\!1\!+\!j\mu(d_m^{(t)}\!-d_n^{(t)})+\!\frac{(j\mu(d_m^{(t)}-d_n^{(t)}))^2}{2}\bigg)\!\Bigg\}
\end{align}
Thus, denoting by $z_{0}$, $z_{1}$, $z_{2}$ the constant, linear, and quadratic term in \eqref{Prob:Stepsizeapp2}, respectively, we obtain the simple approximation
\begin{align} \label{Prob:Stepsizeapp3}
\widehat{h}(\mu)=z_0+z_1\mu-z_2\mu^2\;,
\end{align}
which admits the closed-form maximizer $\mu^{*}=z_{1}/(2z_{2})$ if\footnote{Traditional line search can  be used if these conditions do not hold.} $z_{1}\geq 0$, $z_{2}>0$.

At this point, equipped with an expression for the gradient of the objective function $\mathcal{F}(\mathbf{\Phi}(\mathbf{\Theta}))$ as well as a simple way to obtain the step-size, we employ the Polak-Ribiere-Polyak conjugate gradient algorithm to update the descent direction using the following formula \cite{TANGliang,Stoica2}:
\begin{align}
\mathbf{d}^{(t+1)}=-\mathbf{q}^{(t+1)}+\frac{(\mathbf{q}^{(t+1)}-\mathbf{q}^{(t)})^{T}\mathbf{q}^{(t+1)}}{\|\mathbf{q}^{(t)}\|^2}\mathbf{d}^{(t)},
\end{align}
where $\mathbf{q}^{(t)}\triangleq\nabla_{\mathbf{\Theta}}\left((\mathbf{y}^{(t)})^H\mathbf{A}\mathbf{y}^{(t)}\right)$. It should be mentioned that, in case the step size is computed based on the approximate function $\widehat{h}(\mu)$ in \eqref{Prob:Stepsizeapp3}, it not guaranteed that $\mathbf{d}^{(t)}$ is a descent direction. Thus, in order to guarantee that at each iteration $t$ of the gradient search we obtain a descent direction,
we consider the following modified direction update rule:
 \[
\mathbf{d}^{(t+1)}=
\begin{dcases}
\mathbf{d}^{(t+1)}, & (\mathbf{q}^{(t+1)})^{T} \mathbf{d}^{(t+1)}<0 \\
-\mathbf{q}^{(t+1)}, & (\mathbf{q}^{(t+1)})^{T} \mathbf{d}^{(t+1)}\geq0
\end{dcases}.
\]

\subsubsection{Sequential Fractional Programming}
By substituting \eqref{Eq:eObjResAllProbPhi2} into \eqref{Eq:ResAllProbPhi1}, the optimization problem with respect to $\mathbf{\Phi}$ for given $\mathbf{P}$ can be rewritten as
\begin{subequations}\label{Eq:ResAllProbMM}
\begin{align}
&\displaystyle \min_{\mathbf{\Phi} }\text{vec}(\mathbf{\Phi}^{-1})^{H}\mathbf{A}\text{vec}(\mathbf{\Phi}^{-1}) \label{Eq:aResAllProbMM}\\
&\;\text{s.t.}\;|\phi_n|=1\;\forall n=1,2,\ldots,N, \label{Eq:bResAllProbMM}
\end{align}
\end{subequations}
As it will be shown in the sequel, the form of the objective in \eqref{Eq:aResAllProbMM} enables us to deal with the non-convex constraint \eqref{Eq:bResAllProbMM}, provided that \eqref{Eq:aResAllProbMM} can be reformulated into a differentiable function. To this end, a convenient approach is to resort to the Sequential Fractional Programming (SFP) optimization method, also known as majorization-minimization, or inner approximation method \cite{ZapNow15,AlessioGEE2017,MM_review,MM_sun,MM_song}.

The basic idea of the SFP method is to tackle a difficult problem by solving a sequence of approximate subproblems. If each approximate problem fulfills some assumptions with respect to the original problem, then the sequence converges and first-order optimality holds upon convergence. In more detail, consider the following general optimization program:
\begin{subequations}\label{Prob:MM}
\begin{align}
&\displaystyle\min_{\mathbf{x}}\bar{f}(\mathbf{x})\\
&\;\text{s.t.}\;\; g_{i}(\mathbf{x})\leq 0\;\forall i=1,2,\ldots,I.
\end{align}
\end{subequations}
Consider also the sequence $\{\mathbf{x}^{(t)}\}$ of feasible points for \eqref{Prob:MM}, and the sequence of minimization problems ${\cal P}_{\mathbf{x}^{(t)}}$ with objectives $f(\mathbf{x}|\mathbf{x}^{(t)})$ and the same constraints as \eqref{Prob:MM}. If for each $\mathbf{x}^{(t)}$ the following conditions are fulfilled:
\begin{enumerate}
\item $f(\mathbf{x}|\mathbf{x}^{(t)})\geq \bar{f}(\mathbf{x})$ for all feasible $\mathbf{x}$,
\item $f(\mathbf{x}^{(t)}|\mathbf{x}^{(t)})=\bar{f}(\mathbf{x}^{(t)})$,
\item $\nabla_{\mathbf{x}}f(\mathbf{x}^{(t)}|\mathbf{x}^{(t)})=\nabla_{\mathbf{x}}\bar{f}(\mathbf{x}^{(t)})$,
\end{enumerate}
then the optimal sequence $\{f(\mathbf{x}^{\star})^{(t)}\}$ of the optimal values of problems ${\cal P}_{\mathbf{x}^{(t)}}$ is monotonically decreasing and converges. Moreover, upon convergence, the first-order optimality properties of problem \eqref{Prob:MM} are satisfied \cite{ZapNow15,AlessioGEE2017,MM_review,MM_sun,MM_song}. Clearly, the usefulness of the SFP method depends on the possibility of determining suitable functions $f(\mathbf{x}|\mathbf{x}^{(t)})$ meeting the above three conditions. In addition, these functions need to be easier than \eqref{Eq:aResAllProbMM} to minimize.

For the optimization problem \eqref{Eq:ResAllProbMM} at hand, the SFP method can be applied as described in the rest of this section. We commence with the following lemma that provides a convenient upper bound for the objective $\mathbf{y}^H\mathbf{A}\mathbf{y}$ in \eqref{Eq:aResAllProbMM} (the definition $\mathbf{y}=\mathrm{vec}(\mathbf{\Phi}^{-1})$ in \eqref{eq:y} is reused).
\begin{lemma}\label{Lem:1}
For any feasible $\mathbf{y}$ for \eqref{Eq:ResAllProbMM}, and given any feasible point $\mathbf{y}^{(t)}$, a suitable upper bound to employ the SFP method is:
\begin{align}\label{eq_5}
&\mathbf{y}^{H}\mathbf{A}\mathbf{y}\leq f(\mathbf{y}|\mathbf{y}^{(t)})=\mathbf{y}^H\mathbf{M}\mathbf{y}\! \nonumber \\
&-\!2\mathrm{Re}(\mathbf{y}^H(\mathbf{M}\!-\!\mathbf{A})\mathbf{y}^{(t)})\!+(\mathbf{y}^{(t)})^H(\mathbf{M}\!-\!\mathbf{A})\mathbf{y}^{(t)},
\end{align}
where $\mathbf{M}\triangleq\lambda_{\rm max}\mathbf{I}_{N^{2}}$ with $\lambda_{\rm max}$ being the maximum eigenvalue of $\mathbf{A}$.
\end{lemma}
\begin{IEEEproof}
Let us consider the following inequality:
\begin{equation}
\|(\mathbf{M}-\mathbf{A})^{1/2}\mathbf{y}-(\mathbf{M}-\mathbf{A})^{1/2}\mathbf{y}^{(t)}\|^{2}\geq0.
\end{equation}
Since the matrix $\mathbf{M}-\mathbf{A}$ is positive semidefinite by construction, elaborating the latter inequality yields
\begin{equation}
\mathbf{y}^{H}(\mathbf{M}\!-\!\mathbf{A})\mathbf{y}\!+(\mathbf{y}^{(t)})^{H}(\mathbf{M}\!-\!\mathbf{A})\mathbf{y}^{(t)}\!-\!2\mathrm{Re}(\mathbf{y}^{H}(\mathbf{M}\!-\!\mathbf{A})\mathbf{y}^{(t)})\geq 0.
\end{equation}
By isolating the term $\mathbf{y}^{H}\mathbf{A}\mathbf{x}$, we obtain the bound in \eqref{eq_5}; this completes the proof.
\end{IEEEproof}

Lemma~\ref{Lem:1} provides a suitable expression for the surrogate function in terms of the variable $\mathbf{y}$ to be used with the SFP method. The next step is to reformulate constraint \eqref{Eq:bResAllProbMM} also in terms of the variable $\mathbf{y}$. Here, it should be paid attention to the fact that \eqref{Eq:bResAllProbMM} enforces the diagonal elements of $\mathbf{\Phi}$ to have unit modulus, whereas $\mathbf{y}$ contains the elements of the vectorized $\mathbf{\Phi}$. Thus, only some elements of $\mathbf{y}$ need to have unit modulus, while all others are bound to be zero. More precisely, the $\mathbf{y}$'s elements that must have unit modulus are those with indices of the form $(i-1)N+i$, with $i=1,2,\ldots,N$. For example, if $N=5$, we have the indices $\{1,N+2,2N+3,3N+4,4N+5=N^{2}=25\}$.

Putting all above together, each iteration of the SFP method requires solving the following problem with respect to the variable $\mathbf{y}$ (from which the optimal $\mathbf{\Phi}$ can be easily obtained):
\begin{subequations}\label{Eq:theta3} \vspace{-3mm}
\begin{align}
& \min_{\mathbf{y}}\;  f(\mathbf{y}|\mathbf{y}^{(t)})\label{Eq:theta3a}\\
&\; \text {s.t. } \; |y_{i}|=1\;,\forall i=(n-1)N+n,\;n=1,2,\ldots,N,\label{Eq:theta3b}\\
&\; \;\;\;\quad \; y_{i}=0\;,\forall i\neq (n-1)N+n\;,\;n=1,2,\ldots,N,\label{Eq:theta3c}
\end{align}
\end{subequations}
where $y_{n}$ denotes the $n$-element of vector $\mathbf{y}$. In the following lemma we present the solution for \eqref{Eq:theta3} .

\begin{lemma}
For any $\mathbf{y}^{(t)}$, the optimization problem \eqref{Eq:theta3} is solved by
\begin{equation}\label{Eq:SolutionMM}
y_{i}=\left\{\begin{array}{cc}
\!\!\!\!e^{j\mathrm{arg}(c_{i})}\;,\forall i=(n-1)N+n,\,n=1,2,\ldots,N\\
\!\!0\;,\;\;\;\;\;\;\;\;\;\;\;\forall i\neq (n-1)N+n,\,n=1,2,\ldots,N
\end{array}\right.,
\end{equation}
where $c_{i}$ with $i=1,2,\ldots,N^2$ is the $i$-th element of $\mathbf{c}\triangleq(\lambda_{\rm max}\mathbf{I}_{N^{2}}-\mathbf{A})\mathbf{y}^{(t)}$.
\end{lemma}
\begin{IEEEproof}
The objective function \eqref{Eq:theta3a} can be expressed as
\begin{align}
f(\mathbf{y}|\mathbf{y}^{(t)})&=\lambda_{\rm max}\|\mathbf{y}\|^{2}-2\mathrm{Re}(\mathbf{y}^H(\lambda_{\rm max}\mathbf{I}_{N^{2}}-\mathbf{A})\mathbf{y}^{(t)}) \nonumber \\
&\,\,\,\,\,\,+(\mathbf{y}^{(t)})^H(\lambda_{\rm max}\mathbf{I}_{N^{2}}-\mathbf{A})\mathbf{y}^{(t)}\label{eq_13}.
\end{align}
By neglecting the terms that do not depend on $\mathbf{y}$ and observing that by virtue of \eqref{Eq:theta3b} and \eqref{Eq:theta3c} it holds $\|\mathbf{y}\|^{2}=N$, problem \eqref{Eq:theta3} can be equivalently recast as
\begin{subequations}\label{Eq:theta4}
\begin{align}
& \max_{\mathbf{y}}\;  2\mathrm{Re}(\mathbf{y}^H(\lambda_{\rm max}\mathbf{I}_{N^{2}}-\mathbf{A})\mathbf{y}^{(t)})\label{Eq:theta4a}\\
&\; \text {s.t. } \; |y_{i}|=1\;,\forall i=(j-1)N+j,\;j=1,2,\ldots,N,\label{Eq:theta4b}\\
&\; \;\;\;\quad \; y_{i}=0\;,\forall i\neq (j-1)N+j\;,\;j=1,2,\ldots,N.\label{Eq:theta4c}
\end{align}
\end{subequations}
Clearly, the only free variables are the phases of the $y_{i}$'s components whose modulus is constrained to be unity, whereas all other components of $\mathbf{y}$ are constrained to be zero. Then, \eqref{Eq:theta4a} can be seen to be maximized when the phases of the non-zero components of $y_{i}$'s are aligned with those of the corresponding components of the vector $(\lambda_{\rm max}\mathbf{I}_{N^{2}}-\mathbf{A})\mathbf{y}^{(t)}$. Hence, \eqref{Eq:SolutionMM} is obtained.
\end{IEEEproof}

\subsection{Optimization with respect to the Power Allocation $\mathbf{P}$}\label{Sec:PowerAll}
We now turn again our attention in problem \eqref{Prob:ResAllpower} for the case where $\mathbf{\Phi}$ is fixed and the objective is the optimization over $\mathbf{P}$. Particularly, we focus on solving:
\begin{subequations}\label{Prob:fixedthetaEE}
\begin{align}
&\displaystyle \max_{\mathbf{P}}\;\frac{\sum_{k=1}^{K}\log_2\left(1+p_k\sigma^{-2}\right)}{\xi \sum_{k=1}^{K} p_k + P_{\rm BS} + KP_{\rm UE} + NP_{n}(b)}\label{Prob:afixedthetaEE} \\
&\;\text {s.t. } \;p_{k}\geq \sigma^{2}(2^{R_{{\rm min},k}}-1),\;\forall k=1,2,\ldots,K,\label{Prob:bfixedthetaEE}\\
&\;\quad\;\; \text{tr}((\mathbf{H}_{2}\mathbf{\Phi} \mathbf{H}_{1})^{+}\mathbf{P}(\mathbf{H}_{2}\mathbf{\Phi} \mathbf{H}_{1})^{+H})\leq P_{\rm max}.\label{Prob:cfixedthetaEE}
\end{align}
\end{subequations}
It is can be that, for fixed $\mathbf{\Phi}$, the numerator of \eqref{Prob:afixedthetaEE} is concave in $\mathbf{P}$, while the denominator of \eqref{Prob:afixedthetaEE} is affine in $\mathbf{P}$. Moreover, both constraints \eqref{Prob:bfixedthetaEE} and \eqref{Prob:cfixedthetaEE} are also affine with respect to $\mathbf{P}$. As a consequence, problem \eqref{Prob:fixedthetaEE} is a single-ratio maximization problem that can be globally solved with limited complexity using Dinkelbach's algorithm \cite{ZapNow15}. This method is summarized in Algorithm~\ref{Dinkelbach}, where $\mathcal{B}\triangleq\{\mathbf{P}=\mathrm{diag}[p_1,p_2,\ldots,p_K]: \eqref{Prob:bfixedthetaEE} \,\& \, \eqref{Prob:cfixedthetaEE}\}$ and $\mathbf{P}^{*}_{i}\triangleq\mathrm{diag}[p^*_{1,i},p^*_{2,i},\ldots,p^*_{K,i}]$ denotes the transmit power allocation solution in Step 3 at each $i$-th (with $i=1,2,\ldots$) algorithmic iteration.
\begin{algorithm}[!t]
\caption{Dinkelbach's Method}\label{Dinkelbach}
\begin{algorithmic}[1]
\State \textbf{Initialization:} $K$, $b$, $\xi$, $P_{\rm BS}$, $P_{\rm UE}$, $P_n(b)$, $\epsilon>0$, and $\lambda_0=0$.
\For{$i=1,2,\ldots$}
\State Solve the concave maximization:
\Statex \hspace{0.58cm}$\mathbf{P}^{*}_{i}=\arg\max\limits_{\mathbf{P}\in\mathcal{B}}\sum_{k=1}^{K}\log_2(1+p_k\sigma^{-2})-\lambda_{i-1}(\xi\sum_{k=1}^{K}p_k+P_{\rm BS}+KP_{\rm UE}+NP_n(b))$.
\State Set $\lambda_{i}=\frac{\sum_{k=1}^{K}\log_2\left(1+p^{\star}_{k,i}\sigma^{-2}\right)}{\xi\sum_{k=1}^{K}p^{\star}_{k,i}+P_{\rm BS}+KP_{\rm UE}+NP_n(b)}$.
    \If{$|\lambda_{i}-\lambda_{i-1}|<\epsilon$}
			\State \textbf{Output:} $\mathbf{P}^{*}_{i}$.
		\EndIf
\EndFor
\end{algorithmic}
\end{algorithm}

Putting together the solutions for $\mathbf{\Phi}$ and $\mathbf{P}$ presented respectively in Sections~\ref{sec:respect_Phi} and~\ref{Sec:PowerAll}, our two proposed EE maximization algorithms for the considered RIS-based multi-user MISO system are summarized in Algorithms~\ref{EEmax-grad} and~\ref{EEmax-MM}. As shown, the solutions for $\mathbf{\Phi}$ and $\mathbf{P}$ are alternatively and iteratively updated till reaching convergence. Specifically, since at each iteration $\ell$, both approaches increase the EE value, i.e. $\text{EE}^{(\ell+1)}\geq \text{EE}^{(\ell)}$ for all $\ell$, convergence of the algorithm in the value of the objective is guaranteed. Indeed, the objective is upper-bounded over the feasible set of \eqref{Prob:ResAllpower}, and thus can not increase indefinitely  \cite{TANGliang,Conjgrade,Conjgrade1,Conjgrade2,Stoica2,chongwenICC2019,Aldayel,Palomar2}. However, no global optimality claim can be made due to the following facts: \textit{i}) problem \eqref{Prob:ResAllpower} is not jointly convex  with respect to both $\mathbf{P}$ and $\mathbf{\Phi}$; and \textit{ii}) the proposed methods for optimizing $\mathbf{\Phi}$ (when $\mathbf{P}$ is fixed) are not guaranteed to yield the globally optimal phase matrix.
\begin{algorithm}[!t]
\caption{Gradient-based EE Maximization Algorithm}\label{EEmax-grad}
\begin{algorithmic}[1]
\State \textbf{Input}: $K$, $b$, $\xi$, $P_{\rm BS}$, $P_{\rm UE}$, $P_n(b)$, $P_{\rm max}$, $\sigma^2$, $\{R_{{\rm min},k}\}_{k=1}^{K}$, $\mathbf{H}_{2}$, $\mathbf{H}_{1}$, and $\epsilon>0$.
\State \textbf{Initialization}: $\mathbf{P}^{0}=\frac{P_{\rm max}}{K}\mathbf{I}_K$, $\mathbf{\Phi}^0=\frac{\pi}{2}\mathbf{I}_N$, $\mathbf{q}^0=\nabla_{\mathbf{\Theta}}((\mathbf{y}^0)^H\mathbf{A}\mathbf{y}^0)$, and $\mathbf{d}^0=-\mathbf{q}^0$.
\While{$|\text{EE}^{(\ell+1)}-\text{EE}^{(\ell)}|^2>\epsilon$,}
\State Given $\mathbf{P}$ update $\mathbf{\Phi}$:
\For{$t=0,1,2,\ldots$}
\State   \!\!\!\!\!\!\!\!\! $\mathbf{y}_1^{(t)}\!=\!\mathbf{y}^{(t)}\! \circ \! \mathbf{d}^{(t)}$.
\State   \!\!\!\!\!\!\!\!\! $\mathbf{y}_2^{(t)}\!=\!\mathbf{y}^{(t)} \!\circ\! \mathbf{d}^{(t)}\!\circ \!\mathbf{d}^{(t)}$.
\State\!\!\!\!\!\!\!$z_2=\mathrm{Im}((\mathbf{y}^{(t)})^H\mathbf{A}\mathbf{y}_1^{(t)})$, \State\!\!\!\!\!\!\!$z_3=\mathrm{Re}((\mathbf{y}^{(t)})^H\mathbf{A}\mathbf{y}_2^{(t)})-(\mathbf{y}_1^{(t)})^H\mathbf{A}\mathbf{y}_1^{(t)}$.
\State\!\!\!\!\!\!\!Compute $\widehat{h}(\mu)=0$ using \eqref{Prob:Stepsizeapp3}.
\State\!\!\!\!\!\!\!Set the step size as $\mu=-\frac{z_2}{z_3}$.
\State\!\!\!\!\!\!\!$\mathbf{y}^{(t+1)}=\mathbf{y}^{(t)}\circ e^{j\mu\mathbf{d}^{(t)}} \circ  \text{vec}(\mathbf{I}_N)$.
\State\!\!\!\!\!\!\!$\mathbf{q}^{(t+1)}=2\mathrm{Re}\!\left(\!-j(\mathbf{y}^{\ast})^{(t+1)}\!\circ\! (\mathbf{A}\mathbf{y}^{(t+1)})\right)$.
\State \!\!\!\!\!\!\!$\mathbf{d}^{(t+1)}=-\mathbf{q}^{(t+1)}+\frac{(\mathbf{q}^{(t+1)}-\mathbf{q}^{(t)})^{T}\mathbf{q}^{(t+1)}}{\|\mathbf{q}^{(t)}\|^2}\mathbf{d}^{(t)}$.
\State\!\!\!\!\!\!\!$
\mathbf{d}^{(t+1)}=
\begin{cases}
\mathbf{d}^{(t+1)}, & (\mathbf{q}^{(t+1)})^{T} \mathbf{d}^{(t+1)}<0 \\
-\mathbf{q}^{(t+1)}, & (\mathbf{q}^{(t+1)})^{T} \mathbf{d}^{(t+1)}\geq0
\end{cases}$.
\State\!\!\!\!\!\!\!\textbf{Until} $\|\mathbf{\Phi}^{(t+1)}-\mathbf{\Phi}^{(t)}\|^2<\epsilon$; Obtain $\mathbf{\Phi}^{(\ell+1)}=\mathbf{\Phi}^{(t+1)}$.
\EndFor
\Statex Given $\mathbf{\Phi}$ update $\mathbf{P}$:
\If{\eqref{Eq:aResAllProbPhi1} evaluated at ${\mathbf{\Phi}}^{(\ell+1)}$ is lower than $P_{\rm max}$}:
\State \!\!\!\!\!\!\!\!\!\!\! Update $\mathbf{P}$ by solving the following problem using Algorithm~\ref{Dinkelbach}:
\State  $\mathbf{P}^{(\ell+1)}=\arg\max \limits_{\mathbf{P}\in \mathcal{B}}\;\frac{\sum_{k=1}^{K}\log_2\left(1+p_k\sigma^{-2}\right)}{\xi\sum_{k=1}^{K}p_k + P_{\rm BS} + KP_{\rm UE} + NP_{n}(b)}$ \label{EEDinkpro}
\Else Break and declare infeasibility.
\EndIf
\EndWhile
 \State \textbf{Output:} $\mathbf{\Phi}$ and $\mathbf{P}$.
\end{algorithmic}
\end{algorithm}

\begin{algorithm}[!t]
\caption{Sequential programming EE Maximization Algorithm}\label{EEmax-MM}
\begin{algorithmic}[1]
\State \textbf{Input}: $K$, $b$, $\xi$, $P_{\rm BS}$, $P_{\rm UE}$, $P_n(b)$, $P_{\rm max}$, $\sigma^2$, $\{R_{{\rm min},k}\}_{k=1}^{K}$, $\mathbf{H}_{2}$, $\mathbf{H}_{1}$, and $\epsilon> 0$.
\State \textbf{Initialization}: $\mathbf{P}^{0}=\frac{P_{\rm max}}{K}\mathbf{I}_K$ and $\mathbf{\Phi}^0=\frac{\pi}{2}\mathbf{I}_N$.
\While{$|\text{EE}^{(\ell+1)}-\text{EE}^{(\ell)}|^2>\epsilon$,}
\Statex Optimize with respect to $\mathbf{\Phi}$ given $\mathbf{P}$:
\For{$t=0,1,2,\ldots$}
\State $\mathbf{A}=(\mathbf{\overline{H}}_{2}^{+H}\!\otimes\mathbf{H}_{1}^{+})^H(\mathbf{\overline{H}}_{2}^{+H}\otimes\mathbf{H}_{1}^{+})$.
\State $\mathbf{y} \triangleq\mathrm{vec}(\mathbf{\Phi}^{-1}).$
\State Compute $\mathbf{y}$ as in \eqref{Eq:SolutionMM}.
\State $\mathbf{y}^{(t+1)}=\text{reshape}(\mathbf{y})$;
\State\!\!\!\!\!\!\!\textbf{Until} $\|\mathbf{\Phi}^{(t+1)}-\mathbf{\Phi}^{(t)}\|^2<\epsilon$,
\State Obtain $\mathbf{\Phi}^{(\ell+1)}=\mathbf{\Phi}^{(t+1)}$;
\EndFor
\Statex Optimize with respect to $\mathbf{P}$ given $\mathbf{\Phi}$:
\If{\eqref{Eq:aResAllProbPhi1} evaluated at ${\mathbf{\Phi}}^{(\ell+1)}$ is lower than $P_{\rm max}$}:
\State \!\!\!\!\!\!\!\!\!\!\! Update $\mathbf{P}$ by solving the following problem using Algorithm~\ref{Dinkelbach}:
\State  $\mathbf{P}^{(\ell+1)}=\arg\max\limits_{\mathbf{P}\in \mathcal{B}}\;\frac{\sum_{k=1}^{K}\log_2\left(1+p_k\sigma^{-2}\right)}{\xi\sum_{k=1}^{K}p_k + P_{\rm BS} + KP_{\rm UE} + NP_{n}(b)}$ \label{EEDinkpro}
\Else \quad Break and declare infeasibility.
\EndIf
\EndWhile
 \State \textbf{Output:} $\mathbf{\Phi}^{(\ell+1)}$ and $\mathbf{P}^{(\ell+1)}$.
\end{algorithmic}
\end{algorithm}

\subsection{Sum Rate Maximization}\label{sec:SumRatemax}
In the previous sections we focused on the EE maximization problem \eqref{Prob:ResAllpower}. It should be stressed, however, that sum-rate maximization is a special case of the considered EE maximization problem. More specifically, it can be seen from \eqref{Prob:ResAllpower} that the system sum rate performance is given by the numerator in \eqref{Prob:aResAllpower}. Hence, Algorithm~\ref{EEmax-grad} can be specialized to perform sum rate maximization by simply setting $\xi=0$, since in this case the denominator of \eqref{Prob:aResAllpower} reduces to a constant. In particular, this modification affects the optimization with respect to the transmit powers, which reduces to a non-fractional and convex problem that can be solved by a single iteration of Algorithm~\ref{Dinkelbach}.

\subsection{Computational Complexity}
The computational complexity of the proposed algorithms depends on the number of iterations of the alternating maximization, say $I_{alt}$, and on the complexity required to solve each sub-problem.

As for Algorithm 2, it can be seen that the optimization of the RIS phase shifts depends on the number of gradient updates, say $I_{gd}$, times the amount of operations performed in each gradient update, which can be seen to scale with\footnote{The complexity of additions is negligible compared to that required for the multiplications.} $N^{2}$ complex multiplications. On the other hand, the optimization of the transmit powers requires to perform $I_{D}$ Dinkelbach's iterations, each requiring to solve a convex problem. Thus, recalling that convex problems have a polynomial complexity in the number of optimization variables, which is at most quartic in the number of variables \cite{ben2001lectures}, the asymptotic complexity of Algorithm 2 can be evaluated as
\begin{equation}
\mathcal{O}(I_{alt}(I_{gd}N^{2}+I_{D}K^{p}))\;,
\end{equation}
with $1\leq p\leq 4$. As for the values of $I_{alt}$, $I_{gd}$, and $I_{D}$, deriving closed-form expressions as a function of the system parameters appears prohibitive. Nevertheless, our numerical results confirm that convergence occurs in a few iterations. Moreover, as for $I_{D}$, it is known that Dinkelbach's algorithm exhibits a super-linear convergence rate \cite{ZapNow15}.

A similar analysis holds for Algorithm 3, with the only difference that the complexity of optimizing the RIS phase shifts is given by $I_{seq}N^{q}$, with $1\leq q\leq 4$, since phase optimization performs  $I_{seq}$ iterations of the sequential optimization method, each requiring the solution of a convex problem with $N$ variables. Thus, the asymptotic complexity of Algorithm 3 can be evaluated as
\begin{equation}
\mathcal{O}(I_{alt}(I_{seq}N^{q}+I_{D}K^{p}))\;.
\end{equation}

\section{Numerical Results}\label{Sec:Numerics}
\begin{figure} \vspace{-2mm}
  \begin{center}
  \includegraphics[width=95mm]{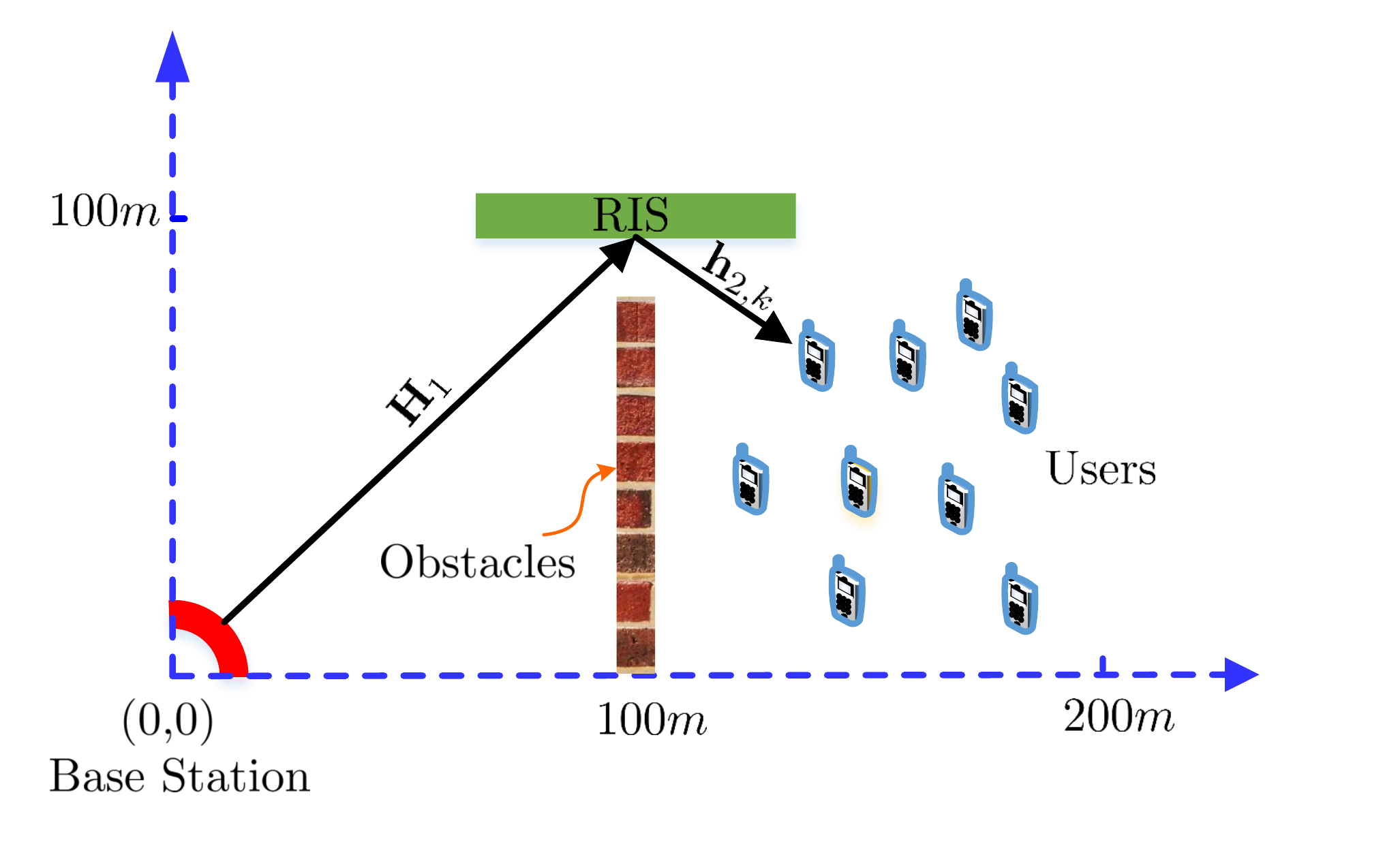}  \vspace{-4mm} %
  \caption{The simulated RIS-based $K$-user MISO communication scenario comprising of a $M$-antenna base station and a $N$-element intelligent surface.  }
  \label{fig:PIM} \vspace{-4mm}
  \end{center}
\end{figure}
\begin{table*}[t] \vspace{-2mm}
\caption{Simulation and Algorithmic Parameters} \label{tabpar} \vspace{-2mm}
\begin{center}
    \begin{tabular}{| l | l || l | l |}
    \hline
    \textbf{Parameters} & \textbf{Values} & \textbf{Parameters} & \textbf{Values} \\ \hline
    RIS central element placement: & $(100m,100m)$ & Circuit dissipated power at BS $P_{\rm BS}$: & $9$dBW  \\
    BS central element placement: & $(0,0)$ &  Circuit dissipated power coefficients at BS $\xi$ and AF relay $\xi_{\rm AF}$: & $1.2$  \\
		Small scale fading model $\forall$$i,k,j$: & $[\mathbf{H}_1]_{ij},[\mathbf{h}_{2,k}]_{i}\sim\mathcal{CN}(0,1)$ &     Maximum transmit power at BS and AF relay $P_{\rm max}$=$P_{\rm R,max}$: & $20$dBW \\
		Large scale fading model at distance $d$: & $\frac{10^{-3.53}}{d^{3.76}}$ & Dissipated power at each user $P_{\rm UE}$: & $10$dBm  \\
    Transmission bandwidth ${\rm BW}$: & $180k\mathrm{Hz}$  & Dissipated power at the $n$-th RIS element $P_n(b)$: & $10$dBm  \\
    Algorithmic convergence parameter: & $\epsilon=10^{-3}$  &      Dissipated power at each AF relay transmit-receive antenna $P_R$: & $10$dBm \\
    \hline
    \end{tabular}
\end{center}
\end{table*}

In this section, we investigate the performance of the RIS-based $K$-user MISO communication system illustrated in Fig$.$~\ref{fig:PIM}. The multiple single-antenna mobile users are assumed randomly and uniformly placed in the $100m\times100m$ half right-hand side rectangular of the figure. All presented illustrations have been averaged results over $10^{3}$ independent realizations of the users' positions and channel realizations, generated according to the 3GPP propagation environment described in \cite{emil2015_fading}, whose parameters are summarized in Table~\ref{tabpar}. Therein, $[\mathbf{H}_1]_{ij}$ and $[\mathbf{h}_{2,k}]_{i}$ with $i=1,2,\ldots,N$, $k=1,2,\ldots,K$, and $j=1,2,\ldots,M$ denote the $(i,j)$-th and $i$-th elements of the respective matrices. In the table above, we also include the hardware dissipation parameters of \cite{emil_setting,ZapNow15} for BS, RIS, and the mobile users, as well as for the AF relay that will be used for performance comparisons purposes. The relay is assumed to transmit with maximum power $P_{{\rm R},\rm max}$, which is considered in all performance results equal to $P_{\rm max}$.

Without loss of generality, in the figures that follow we assume equal individual rate constraints for all $K$ users, i$.$e$.$, $R_{{\rm min},k}=R_{\rm min}$ $\forall$$k$. In addition, we set $R_{\rm min}$ to a fraction of the rate that each user would have in the genie case of mutually orthogonal channels and uniform power allocation. In particular, this genie rate for each $k$-th mobile user is given by
\begin{align}\label{NumResults1}
R= \log_2\left(1+\frac{P_{\rm max}}{K\sigma^2}\right).
\end{align}
Thus, the QoS constraints depend on $P_{max}$, which ensures that the minimum rate is commensurate to the maximum power that can be used, in turn leading to a feasibility rate of Problem \eqref{Prob:ResAllpower} that is approximately constant with $P_{max}$. Table \ref{QoS} below shows the feasibility rate obtained for $P_{max}=20\;\textrm{dBW}$ and different fractions $R$.

\begin{table}[ht!] \vspace{-4mm}
\caption{ Different QoSs VS feasibility rate } \label{QoS}
\begin{center}
    \begin{tabular}{| p{3cm} | p{0.6cm} |p{0.6cm}|p{0.6cm}|p{0.6cm}|p{0.6cm}|}
    \hline
    \textbf{QoS} ($\mathrm{bps/Hz}$) & \textbf{0.1}R & \textbf{0.2}R & \textbf{0.3}R& \textbf{0.4}R& \textbf{0.5}R  \\    \hline
      Feasibility Rate (\%) & 99.44 & 99.44&99.44&99.23&99.02\\    \hline
    \end{tabular}
\end{center}  \vspace{-4mm}
\end{table}
Moreover, we mention that, in the very few unfeasible scenarios encountered in our simulation, we have relaxed the QoS constraint and deployed the corresponding solution. In light of the high feasibility rate shown above, this has a negligible impact on the shown results.

\subsection{Benchmark: Amplify-and-Forward Relay} \label{Relay}
\begin{figure} \vspace{-2mm}
  \begin{center}
  \includegraphics[width=84mm]{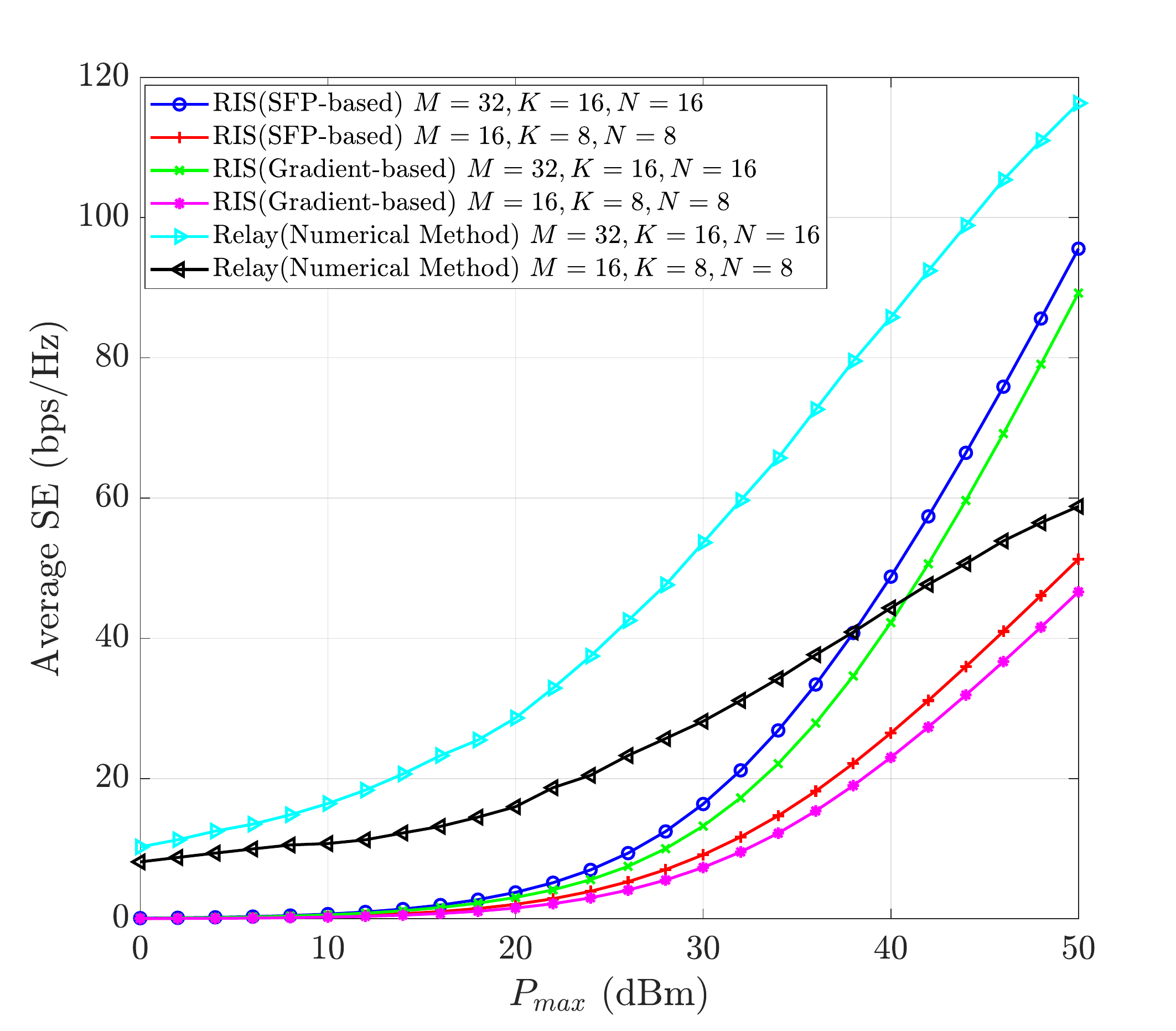}  \vspace{-3mm} %
  \caption{Average SE using either RIS or AF relay versus $P_{\rm max}$ for $R_{\rm min}=0$bps/Hz and: a) $M=32$, $K=16$, $N=16$; and b) $M=16$, $K=8$, $N=8$.}
  \label{fig:ombing} \vspace{-6mm}
  \end{center}
\end{figure}
It is reasonable to expect that the consideration of a reconfigurable RIS structure in the investigated scenario of Fig$.$~\ref{fig:PIM} provides substantial EE gains compared to the case where such a surface is absent; this intuition has been verified via simulations in \cite{chongwen2018}. Hereinafter, we consider a more relevant to Fig$.$~\ref{fig:PIM} benchmark scheme that includes a conventional $N$-antenna AF relay \cite{relay_model,Relay_precoding,relay_heath,relay_allessio} in the place of the RIS structure. To ensure a fair comparison between this benchmark scheme and our proposed RIS-based one in the performance results that follow, we have considered the same users' positions and channel realizations in both cases. Similar to the RIS case modeled by the phase shifting matrix $\mathbf{\Phi}$, we assume that the relay deploys the $N\times N$ complex diagonal AF matrix $\mathbf{V}$. Differently from $\mathbf{\Phi}$, $\mathbf{V}$'s diagonal elements are not constrained to have unit modulus, but rather a maximum relay power constraint is enforced. In more detail, the baseband received signals $\mathbf{y}_{\rm R}\in\mathbb{C}^{N\times1}$ and $\mathbf{y}_K\in\mathbb{C}^{K\times1}$ at the relay and at all $K$ mobile users, respectively, can be expressed as
\begin{align}\label{relay_m1}
   \mathbf{y}_{\rm R}&\triangleq\mathbf{H}_{1}\mathbf{x}+\mathbf{w}_{\rm R}, \\
   \mathbf{y}_K&\triangleq\mathbf{H}_{2}\mathbf{V}\mathbf{y}_{\rm R}+\mathbf{w}_K=\mathbf{H}_{2}\mathbf{V}\mathbf{H}_{1}\mathbf{x}+\mathbf{H}_{2}\mathbf{V}\mathbf{w}_r+\mathbf{w}_K,
\end{align}
where $\mathbf{w}_{\rm R}\in\mathbb{C}^{N\times1}$ denote the thermal noise at relay modeled as a zero-mean complex circularly Gaussian vector with covariance matrix $\mathbf{I}_N$. We model similarly the thermal noises at all $K$ users, which are included in $\mathbf{w}_K\in\mathbb{C}^{K\times1}$ having covariance matrix $\mathbf{I}_K$.

It is interesting to observe that, since the AF matrix $\mathbf{V}$ is not unitary, it introduces a noise amplification effect that is not present in the RIS case. Moreover, as already anticipated, unlike RIS, the AF relay consumes RF power to amplify the incoming signal. Accounting for ZF transmission from BS as in the RIS design case in \eqref{Prob:ResAllpower}, the relay power consumption is given by
\begin{equation}
P_{\rm AF}\triangleq\mathrm{tr}(\mathbf{H}_2^{+}\mathbf{P}\mathbf{H}_2^{+H}+\mathbf{V}\mathbf{V}^{H}\sigma^2).
\end{equation}

Then, for the case of AF relaying, we consider the following EE maximization problem for the joint design of $\mathbf{P}$ and $\mathbf{V}$:
\begin{subequations}\label{Prob:realypro}
\begin{align}
&\displaystyle \max_{\mathbf{V},\mathbf{P}}\; \frac{\sum_{k=1}^{K}\log_2\left(1+\frac{p_k}{ |\mathbf{h}_{2,k}\mathbf{V}\mathbf{V}^{H}\mathbf{h}_{2,k}^{H}|^2+\sigma^2}\right)}{\xi\sum_{k=1}^{K}p_k+P_{\rm BS} + KP_{\rm UE}+\xi_{\rm AF}P_{\rm AF}+NP_{\rm R}}\label{Prob:arealypro}\\
&\;\text{s.t.}\;\log_2\!\left(\!1\!+\!\frac{p_k}{ |\mathbf{h}_{2,k}\mathbf{V}\mathbf{V}^{H}\mathbf{h}_{2,k}^{H}|^2\!+\!\sigma^2}\right)\!\geq \!R_{{\rm min},k}\;\forall k\in\![1,K], \label{Prob:brealypro}\\
&\;\quad\;\; \text{tr}((\mathbf{H}_{2}\mathbf{V} \mathbf{H}_{1})^{+}\mathbf{P}(\mathbf{H}_{2}\mathbf{V} \mathbf{H}_{1})^{+H})\leq P_{\rm max},\label{Prob:crealypro}\\
&\;\quad\;\; \mathrm{tr}(\mathbf{H}_2^{+}\mathbf{P}\mathbf{H}_2^{+H}+\mathbf{V}\mathbf{V}^{H}\sigma^2) \leq P_{{\rm R},\rm max}, \label{Prob:drealypro}
\end{align}
\end{subequations}
where $\xi_{\rm AF}$ depends on the efficiency of the relay power amplifier. In order to solve \eqref{Prob:realypro} we resort again to the alternating optimization method, like in the RIS case.  Moreover, it is assumed that the end-to-end data transmission phase of the relay has the same duration as the end-to-end data  transmission duration of the RIS-based system. Hence, no pre-log factor is considered in \eqref{Prob:realypro}, in analogy with Problem \eqref{Prob:ResAllpower}.

The optimization with respect to $\mathbf{P}$ for a given $\mathbf{V}$ can be performed following a similar approach to Section~\ref{Sec:PowerAll}, whereas the optimization with respect to $\mathbf{V}$ for a given $\mathbf{P}$ becomes more challenging due to the presence of constraint \eqref{Prob:crealypro}. To find the optimum $\mathbf{V}$ in the results that follow we have employed numerical exhaustive search.

\subsection{RIS vs AF relay Performance Comparison}
\begin{figure}\vspace{-2mm}
  \begin{center}
  \includegraphics[width=86mm]{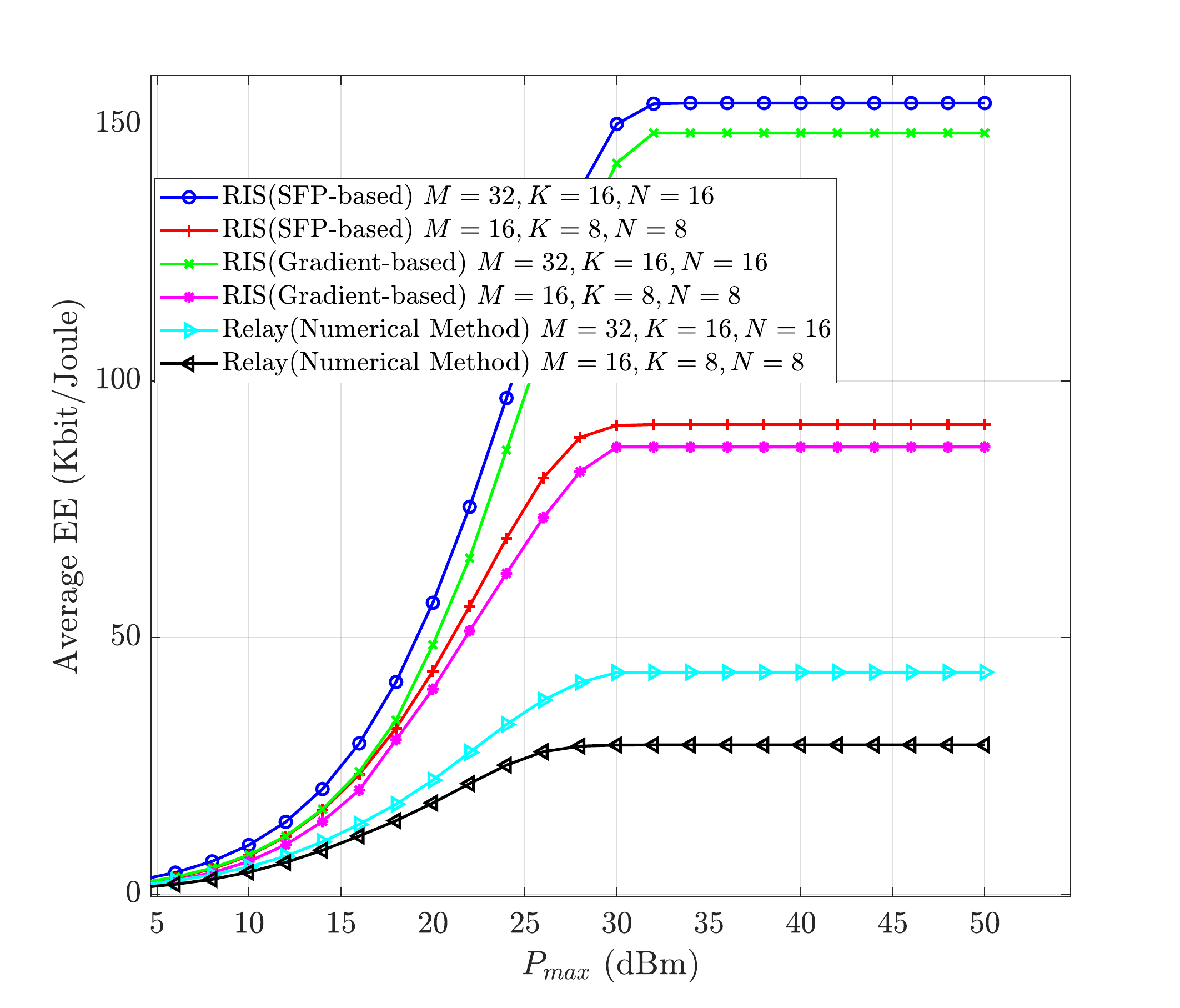}  \vspace{-3mm} %
  \caption{Average EE using either RIS or AF relay versus $P_{\rm max}$ for $R_{\rm min}=0$bps/Hz and: a) $M=32$, $K=16$, $N=16$; and b) $M=16$, $K=8$, $N=8$.}
  \label{fig:EEmax_EE_KMN_PIM}
  \end{center}\vspace{-6mm}
\end{figure}
The achievable SE and EE performances as functions of $P_{\rm max}$ in dBm are illustrated in Figs$.$~\ref{fig:ombing} and~\ref{fig:EEmax_EE_KMN_PIM}, respectively. We have evaluated the gradient- and SFP-based approaches described in Algorithms~\ref{EEmax-grad} and~\ref{EEmax-MM}, respectively, as well as the algorithm described in Section~\ref{Relay} for the AF relay case.
In both figures, we have set the minimum QoS constraint as $R_{\rm min}=0$bps/Hz for all $K$ users, and considered the two different settings: a) $M=32$, $K=16$, and $N=16$; and b)  $M=16$, $K=8$, and $N=8$. As seen from Fig$.$~\ref{fig:ombing}, the relay-assisted system outperforms the RIS-based one, irrespective of the proposed algorithm used. This behavior is expected since the AF relay is an active terminal rather than a reflecting structure as the RIS is. The relay possesses dedicated transmit circuitry that   provides the transmit power $P_{{\rm R},\max}$ to it. Moreover, the relay is not constrained by the unit modulus constraint that the intelligent surface has. Specifically, RIS looses about $40$bps/Hz and $20$bps/Hz at $P_{\rm max}=30$dBm under the settings a) and b), respectively. However, as $P_{\rm max}$ increases, the performance gap between the RIS and relay cases becomes smaller. This is happens because as $P_{\rm max}$ increases, the relay transmit power $P_{{\rm R},\max}$ becomes less and less relevant to the SE, which is actually impacted by the BS transmit power. It can be also observed from Fig$.$~\ref{fig:ombing}, which also holds in Figs$.$~\ref{fig:EEmax_EE_KMN_PIM}, that both proposed Algorithms~\ref{EEmax-grad} and~\ref{EEmax-MM} perform similarly, with the SFP-based one achieving slightly better performance.

The trend in Fig$.$~\ref{fig:ombing} is reversed in Fig$.$~\ref{fig:EEmax_EE_KMN_PIM}, where the EE performance is sketched. It is shown that both proposed algorithms for the RIS-based system case significantly outperform our derived algorithm for the relay-assisted one. Particularly, the EE of the RIS-based system is $300\%$ larger than that of the one based on the AF relay when $P_{\rm max}\geq32$dBm. This is a direct consequence of the fact that former system exhibits a much lower energy consumption compared to the latter one. It is also shown that the setting a) with $M=32$, $K=16$, and $N=16$ is more energy efficient than the setting b) with $M=16$, $K=8$, $N=8$. It can be also observed that for both systems cases the EE performance saturates for $P_{\rm max}\geq32$dBm. This explained by the fact that the EE function is not monotonically increasing with the maximum BS transmit power $P_{\rm max}$, but instead has a finite maximizer. When $P_{\rm max}\geq 32$dBm, the excess BS transmit power is actually not used since as it would only reduce the EE value.

\subsection{Impact of the QoS Constraints}\label{sec:majhead}
\begin{figure}\vspace{-2mm}
  \begin{center}
  \includegraphics[width=86mm]{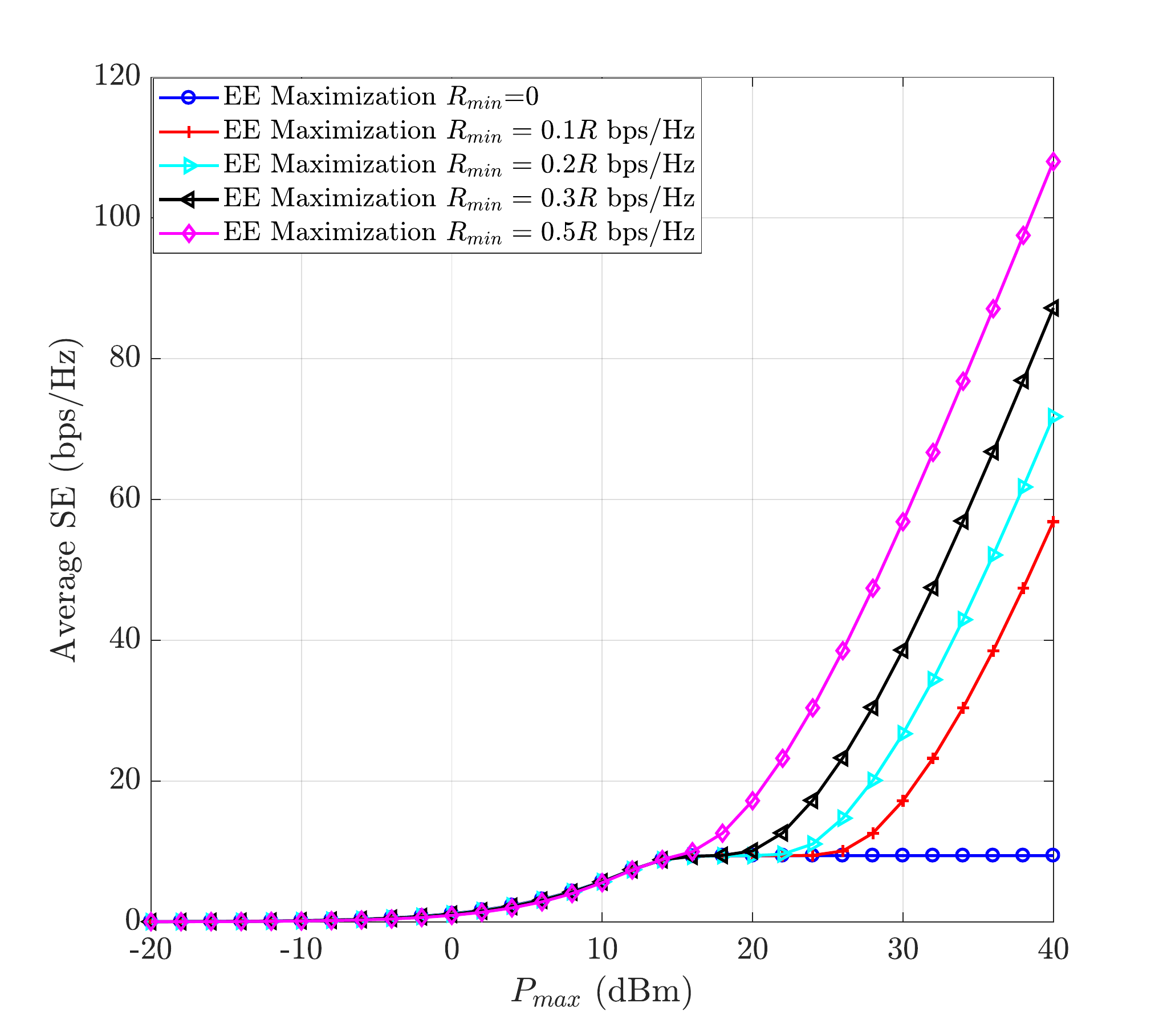}  \vspace{-4mm} %
  \caption{Average SE using RIS versus $P_{\rm max}$ for $M=32$, $K=16$, $N=16$, as well as different fractions of $R$ for $R_{\rm min}$.}
  \label{fig:SE_EEmax_QoSall} \vspace{-6mm}
  \end{center}
\end{figure}
\begin{figure}\vspace{-2mm}
  \begin{center}
  \includegraphics[width=88mm]{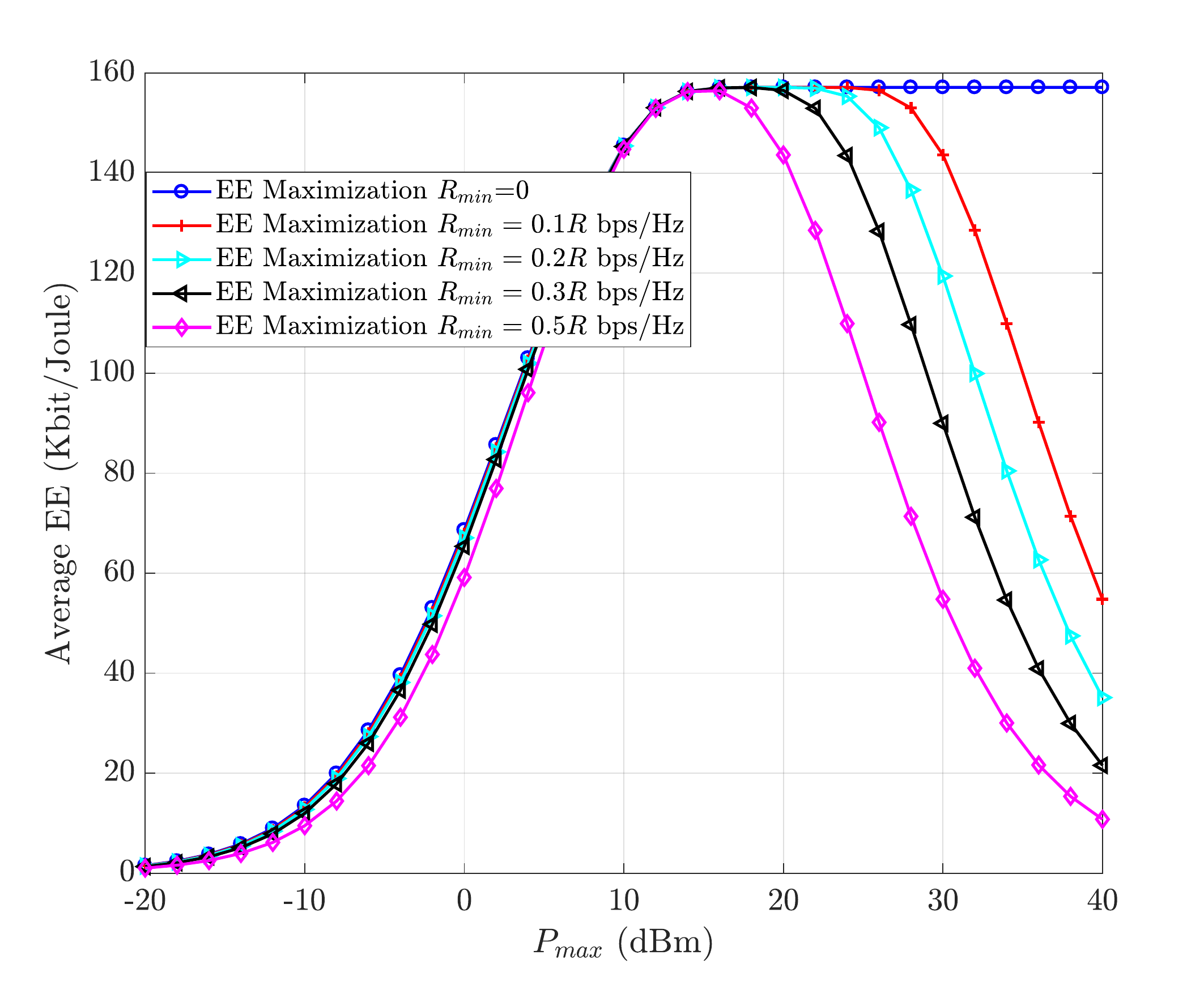}  \vspace{-4mm} %
  \caption{Average EE using RIS versus $P_{\rm max}$ for $M=32$, $K=16$, $N=16$, as well as different fractions of $R$ for $R_{\rm min}$.}
  \label{fig:EEmax_QoSall}
  \end{center}\vspace{-6mm}
\end{figure}

The effect of the different values for $R_{\rm min}$ in the SE and EE performances versus $P_{\rm max}$ in dBm is depicted in Figs$.$~\ref{fig:SE_EEmax_QoSall} and~\ref{fig:EEmax_QoSall}, respectively, using our SFP-based algorithm~\ref{EEmax-MM}. For the cases where the design problems turned out to be infeasible, the rate constraint has been removed and the unconstrained solutions were retained. Fig.~\ref{fig:SE_EEmax_QoSall} shows that for low $P_{\rm max}$ values the problem is nearly always infeasible. This is expected since there is not enough transmit power from BS to meet the rate requests of the users, and thus, the performance of all designed solutions coincide to very low SE values. However, for $P_{\rm max}\geq16$dBm, the values for $R_{\rm min}$ start having a significant impact on the SE. It can be observed that, increasing $R_{\rm min}$ results in increasing the achievable SE and outperforming more the saturating unconstrained case of $R_{\rm min}=0$bps/Hz. Obviously, the larger the $R_{\rm min}$value is, the higher the slope of the SE curve. The performance behavior in Fig$.$~\ref{fig:EEmax_QoSall} follows the same trend with Fig$.$~\ref{fig:SE_EEmax_QoSall}. We may see that for larger $P_{\rm max}$, enforcing stricter QoS constraints causes the EE to decrease faster, due to the fact that the excess BS transmit power is used to meet the common user rate requirements.

\subsection{Comparison between the SE and EE Maximizing Designs}
\begin{figure}\vspace{-2mm}
  \begin{center}
  \includegraphics[width=86mm]{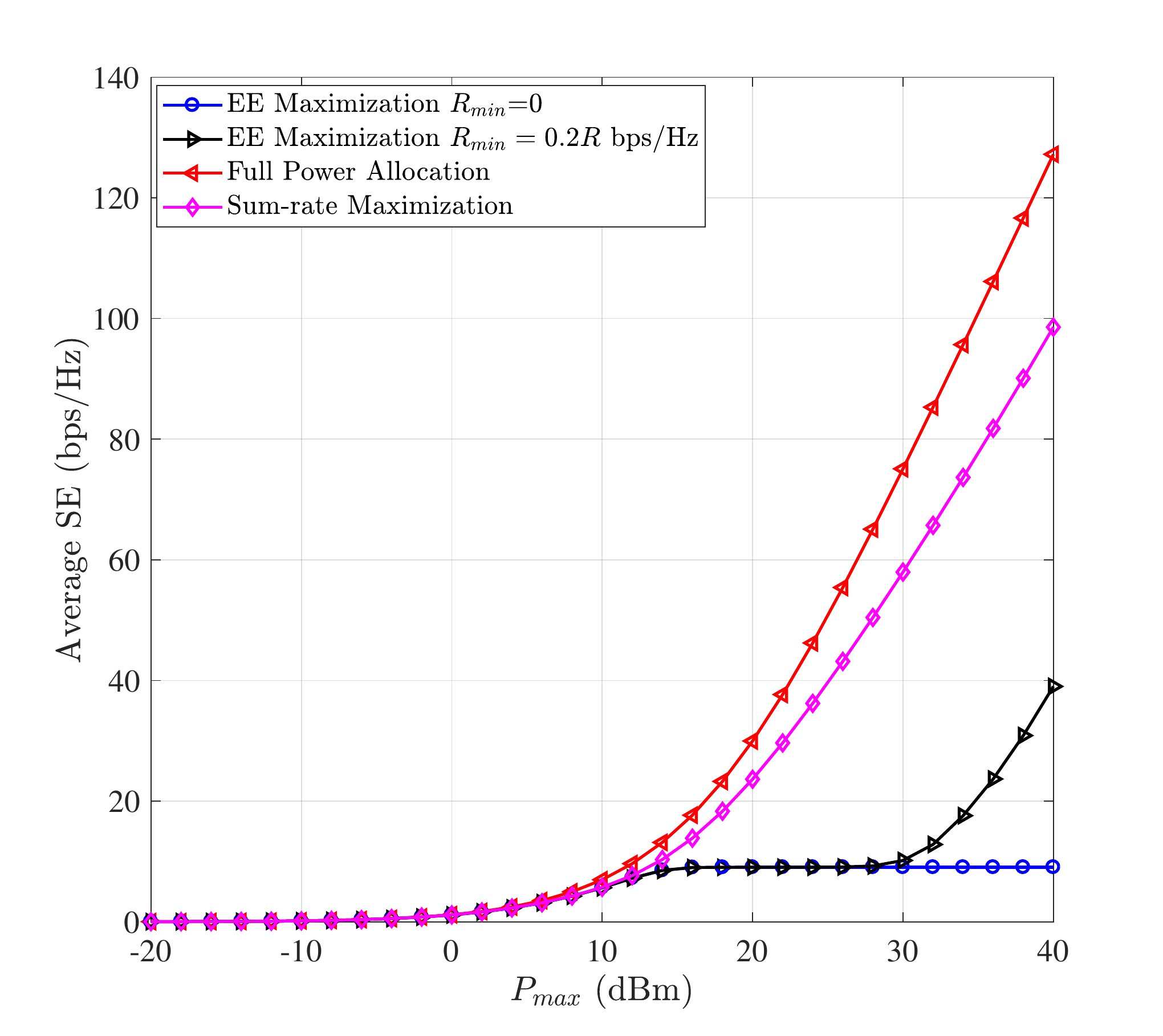}  \vspace{-2mm} %
  \caption{Average SE using RIS versus $P_{\rm max}$ for $M=32$, $K=16$, and $N=16$ using: a) our SFP-based EE maximization algorithm for $R_{\rm min}=\{0,0.2R\}$; b) full power allocation; and c) our SE maximization algorithm.}
  \label{fig:SE_EEmax_QoS_Pmax} \vspace{-6mm}
  \end{center}
\end{figure}

\begin{figure}\vspace{-0mm}
  \begin{center}
  \includegraphics[width=87mm]{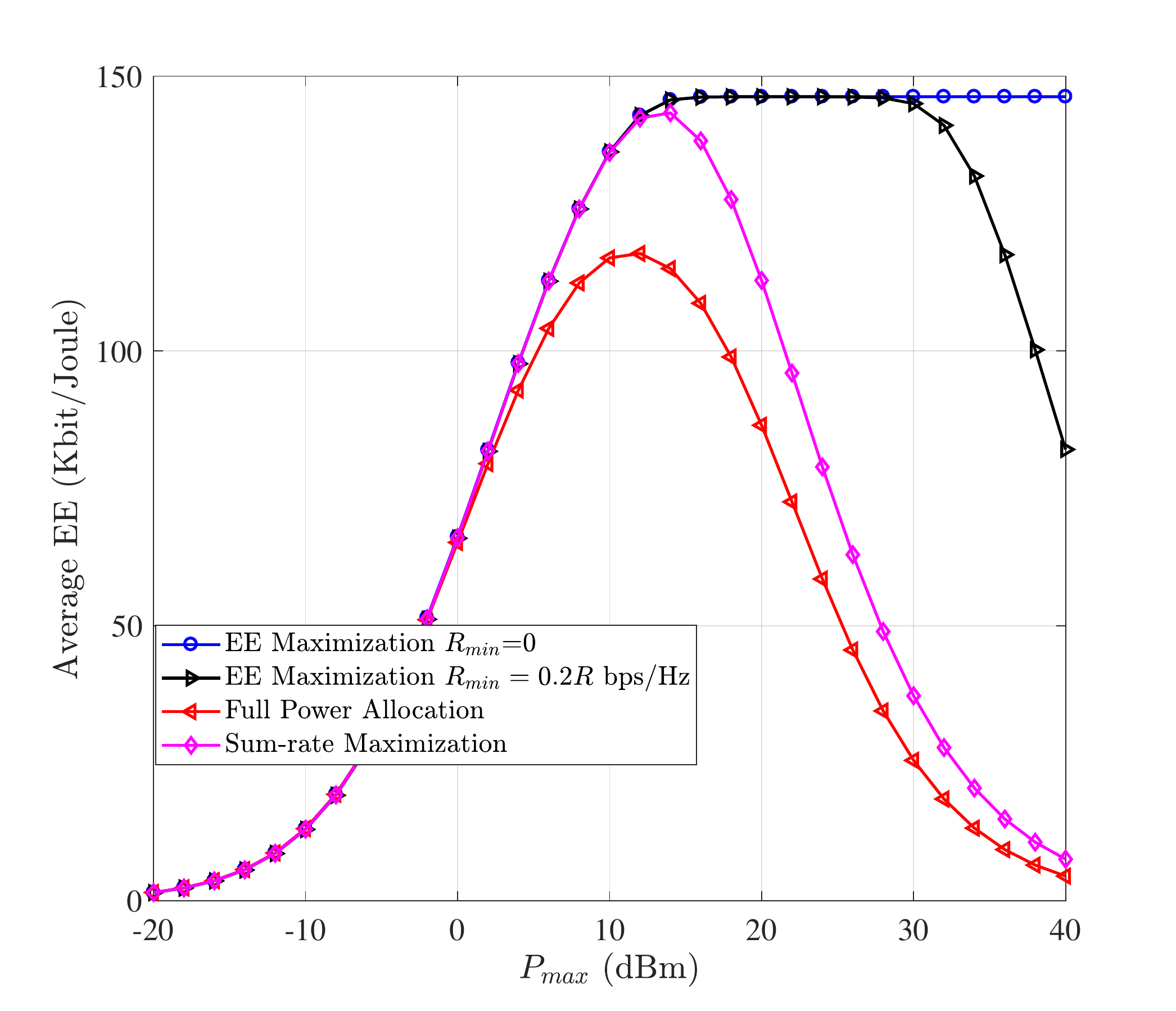}  \vspace{-2mm} %
  \caption{Average EE using RIS versus $P_{\rm max}$ for $M=32$, $K=16$, and $N=16$ using: a) our SFP-based EE maximization algorithm for $R_{\rm min}=\{0,0.2R\}$; b) full power allocation; and c) our SE maximization algorithm.}
  \label{fig:EE_EEmax_QoS_Pmax} \vspace{-6mm}
  \end{center}
\end{figure}

In Figs$.$~\ref{fig:SE_EEmax_QoS_Pmax} and~\ref{fig:EE_EEmax_QoS_Pmax} we plot the average achievable SE and EE performances versus $P_{\rm max}$ in dBm for the following design objectives: \textit{i}) EE maximization with $R_{\rm min}=0$bps/Hz using Algorithm~\ref{EEmax-grad}; \textit{ii}) the same objective and algorithm with \textit{i}) but with $R_{\rm min}=0.2R$bps/Hz; \textit{iii}) SE maximization using the algorithm described in Section~\ref{sec:SumRatemax}; and \textit{iv}) maximum power allocation $P_{\rm max}$ to each user. It can be seen from both figures that all designs perform similarly for $P_{\rm max}\leq15$dBm, which indicates that the EE and SE objectives are nearly equivalent for such transmit power levels. This can be explained by observing that for low $P_{\rm max}$, the EE is an increasing function of the BS transmit power, just as the SE is. In other terms, using full BS transmit power for low $P_{\rm max}$ is optimal, and in this case, EE maximization reduces to SE maximization. However, as shown for $P_{\rm max}>15$dBm, the EE maximizing objective and the SE one result in designs yielding substantially different performances. For such $P_{\rm max}$ values, maximizing the SE requires utilizing all available BS power, whereas maximizing EE does require increasing the BS transmit power above a threshold value. As a result, the SE maximization design naturally increases the SE, but leads to decreasing EE. On the contrary, when maximizing EE both the achievable EE and SE performances become constant. It can be also observed from the results of Figs$.$~\ref{fig:EE_EEmax_QoS_Pmax} and~\ref{fig:SE_EEmax_QoS_Pmax} that, when EE is maximized subject to QoS constraints, an intermediate behavior is obtained due to the fact that some of the excess transmit power is used in order to fulfill the those constraints. Once the constraints are met, no further BS transmit power is needed.
\subsection{Impact of the number of RIS Elements}

\begin{figure}\vspace{-2mm}
  \begin{center}
  \includegraphics[width=85mm]{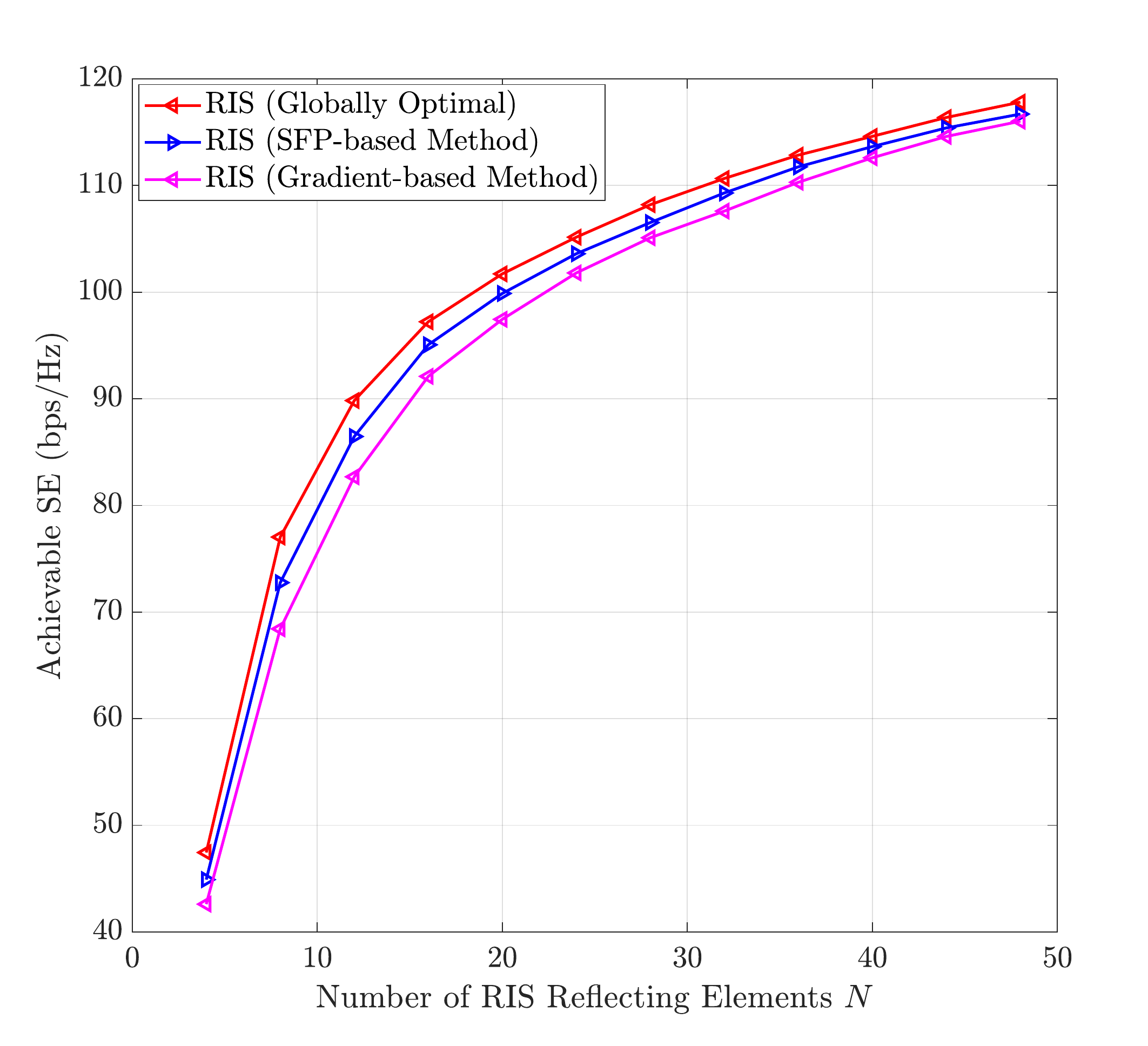}  \vspace{-3mm} %
  \caption{Average SE using RIS versus $N$ for ${\rm SNR}=20$dB, $M=64$, $K=N$ and $R_{\rm min}=2$bps/Hz with both our presented algorithms as well as exhaustive global optimization.}
  \label{fig:N_8_40} \vspace{-6mm}
  \end{center}
\end{figure}

In Fig$.$~\ref{fig:N_8_40} we consider the gradient- and SFP-based Algorithms for  SE maximization, that is a special case of Algorithms~\ref{EEmax-grad} and~\ref{EEmax-MM} respectively, as described in Section~\eqref{sec:SumRatemax}. The SE performance versus the number of the RIS reflecting elements $N$ is shown. We have set the transmit Signal Noise Ratio (SNR), defined as ${\rm SNR}=P_{\max}/\sigma^2$, to $20$dB. Other parameters are set as $M=64$, $K=N$, and $R_{\rm min}=2$bps/Hz.  In this figure, we also report the performance of the optimum SE design obtained by means of numerical global optimization implemented by a numerical grid search. Clearly, this entails an exponential complexity and is considered here only for benchmarking purposes.
\begin{figure}\vspace{-1mm}
  \begin{center}
  \includegraphics[width=83mm]{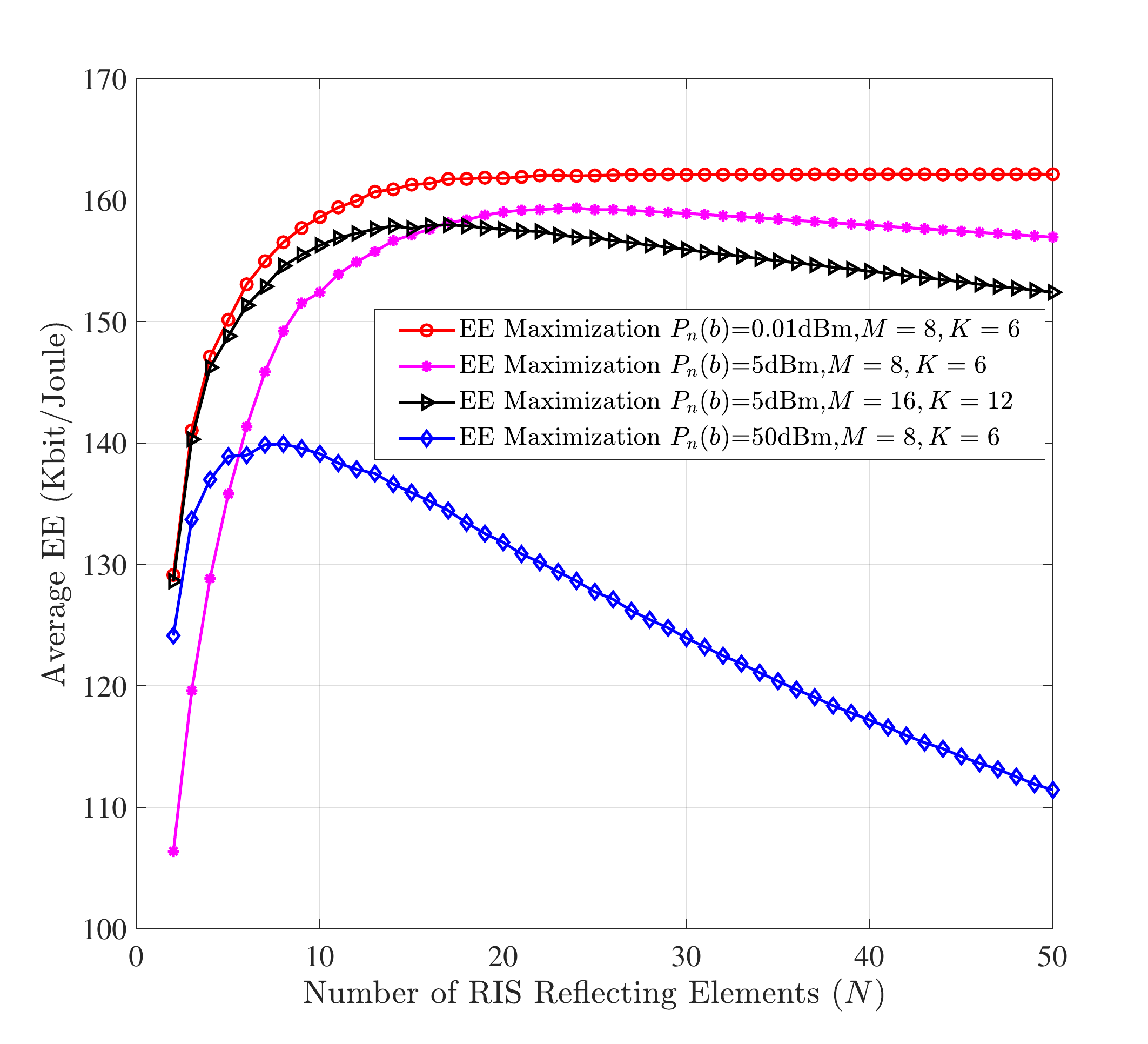}  \vspace{-3.0mm} %
  \caption{Average EE using RIS versus $N$ for ${\rm SNR}=-10$dB and $R_{\rm min}=0$bps/Hz, as well as: a) $P_n(b)=0.01$dBm, $M=8$, and $K=6$; b) $P_n(b)=5$dBm, $M=8$, and $K=6$; c) $P_n(b)=5$dBm, $M=16$, and $K=12$; and d) $P_n(b)=100$dBm, $M=8$, and $K=6$.   Note that the EE performance is implemented by  the numerical grid search method. }
  \label{fig:Com_EE_vs_N} \vspace{-6mm}
  \end{center}
\end{figure}
As clearly shown, both proposed algorithms yield very similar performance curves that are quite close to the ones obtained from the global optimization method. Furthermore, as expected, the larger the $N$ value is, the larger is the achievable SE for the considered RIS-based system.

We finally plot the achieved EE by using the  numerical grid search method in Fig$.$\ref{fig:Com_EE_vs_N} versus $N$. In this figure we consider the following cases for the number of users $K$, the number of BS antenna elements $M$, and the power consumption of each RIS $b$-bit phase resolution element: \textit{i}) $P_n(b)=0.01$dBm, $M=8$, and $K=6$; \textit{ii}) $P_n(b)=5$dBm, $M=8$, and $K=6$; \textit{iii}) $P_n(b)=5$dBm, $M=16$, and $K=12$; and \textit{iv}) $P_n(b)=100$dBm, $M=8$, and $K=6$. As observed, when $N$ is quite small, i$.$e$.$, $N\leq5$, all designs exhibit the same trend. Particularly, EE performance increases as $N$ increases. However, for a small-to-moderate $N$ value and on, EE starts decreasing. This behavior seems not to happen for $P_n(b)=0.01$dBm, but this is only due to the fact that this value of $P_n(b)$ is quite small, and thus it would take a very large $N$ to observe EE decreasing. The results in Fig$.$\ref{fig:Com_EE_vs_N} confirm that there exists an optimal number $N$ of RIS elements as far as the EE maximization objective is concerned. In other words, an optimal trade-off exists between the rate benefit of deploying larger and larger RIS structure and its corresponding energy consumption cost.

\section{Conclusion and Future Work}\label{sec:prior}
In this paper, we considered a RIS-based downlink multi-user MISO system and presented two computationally efficient EE maximization algorithms for the BS transmit power allocation and the RIS reflector values. Both algorithms were based on alternating maximization, with the one adopting gradient descent for the RIS design, while the other is a SFP-based approach. The optimal transmit power allocation was tackled by a fractional programming method. Special cases of both algorithms were used for the SE maximization design. Our numerical results showed that the proposed SFP-based approach  achieves near optimal SE performance, and that RIS-based communication can provide up to $300\%$ higher EE than the relay-assisted one. It was also substantiated that the EE-optimal operating point depends on the numbers of mobile users and RIS elements, as well as the individual power consumption of the RIS elements.

This paper has provided a first contribution in the area of radio resource allocation in cellular communication systems incorporating RIS structures. Many relevant extensions of work can be identified. First of all, this work has considered intelligent surfaces with discrete units, while the investigation of  RIS with reflecting elements coated on the surface in continuity appears as a relevant future research direction. In this context, it becomes relevant to analyze the asymptotic regime in which the number of RIS elements tends to infinity, as well as the study of how EE varies with the area of the surface. Other interesting research directions include the consideration of the direct channel between the BS and the mobile users, the acquisition of channel knowledge when a RIS is included, as well as a thorough study of the trade-off between EE and SE in RIS-assisted communication systems. The impact of channel estimation and feedback overhead on the performance of RIS-based systems needs to be analyzed in detail, too. Finally, another future interesting line of research is the investigation of the impact of the area of RIS on the performance. In \cite{sha_hu2,sha_hu}, it was shown that dividing a given surface area into multiple, smaller, RIS provides more robustness, at the cost of higher feedback and complexity overheads. Moreover, a larger area allows equipping more reflectors and thus provides more optimization variables to tune. However, on the other hand, deploying more reflectors leads to a larger hardware power consumption of the RIS, since the overall power consumption of the RIS increases with the number of equipped reflecting elements. The trade-offs between all these factors need to be better understood, and the optimal value of the surface area that maximizes the system energy efficiency performance needs to be determined.

\section*{Acknowledgement}
The work of C. Yuen was supported by the MIT-SUTD International design center and NSFC 61750110529 Grant, and that of C. Huang by the PHC Merlion PhD program. The work of A. Zappone and M. Debbah was supported by H2020 MSCA IF BESMART, Grant 749336. The work of M. Debbah was also supported by the H2020-ERC PoC-CacheMire, Grant 727682.
\bibliographystyle{IEEEbib}
\bibliography{ris_reference_twc}

\end{document}